\documentclass[11pt]{article}
\usepackage[left=1.5in, right=1in, top=1in, bottom=1in]{geometry}
\usepackage{url}
\usepackage[numbers, square, comma, sort]{natbib}
\usepackage{amsmath,amsthm}
\usepackage{amssymb}
\usepackage{amsfonts}
\usepackage{csquotes}
\usepackage{graphicx, caption,array}
\usepackage[colorlinks,linkcolor={blue},urlcolor ={blue},citecolor={blue}]{hyperref}

\newtheorem{thm}{Theorem}[section]

\theoremstyle{definition}

\theoremstyle{remark}
\newtheorem{rem}[thm]{Remark}
\numberwithin{equation}{section}

\begin{document}

\title{Estimating Robot Strengths with Application to Selection of Alliance Members in FIRST Robotics Competitions}

\author{Alejandro Lim\footnote{University of California, Berkeley, USA, allim@berkeley.edu} \hspace{0.25in} Chin{-}Tsang Chiang\footnote{Institute of Applied Mathematical Sciences, 
National Taiwan University, Taipei, Taiwan, ROC, chiangct@ntu.edu.tw; corresponding author} \hspace{0.25in} Jen-Chieh Teng\footnote{ Institute of Applied Mathematical Sciences,
National Taiwan University, Taipei, Taiwan, ROC, jcteng@ntu.edu.tw}}

\date{December 31, 2020}%

\maketitle

\begin{abstract}
Since the inception of the FIRST Robotics Competition (FRC) and its special playoff system, robotics teams have longed to appropriately quantify the strengths of their designed robots. The FRC includes a playground draft-like phase (alliance selection), arguably the most game-changing part of the competition, in which the top-8 robotics teams in a tournament based on the FRC's ranking system assess potential alliance members for the opportunity of partnering in a playoff stage. In such a three-versus-three competition, several measures and models have been used to characterize actual or relative robot strengths. 
However, existing models are found to have poor predictive performance due to their imprecise estimates of robot strengths caused by
a small ratio of the number of observations to the number of robots. A more general regression model with latent clusters of robot strengths is, thus, proposed to enhance their predictive capacities. Two effective estimation procedures are further developed to simultaneously estimate the number of clusters, clusters of robots, and robot strengths.
Meanwhile, some measures are used to assess the predictive ability of competing models, the agreement between published FRC measures of strength and model-based robot strengths of all, playoff, and FRC top-8 robots, and the agreement between FRC top-8 robots and model-based top robots. Moreover, the stability of estimated robot strengths and accuracies is investigated to determine whether the scheduled matches are excessive or insufficient. In the analysis of qualification data 
from the 2018 FRC Houston and Detroit championships, the predictive ability of our model is also shown to be significantly better than those of existing models. Teams who adopt the new model can now appropriately rank their preferences for playoff alliance partners with greater predictive capability than before.
\end{abstract}


\maketitle

\section{Introduction} \label{sec:intro} \nocite{*}

For Inspiration and Recognition of Science and Technology (FIRST) is an international youth organization which sponsors the FIRST Robotics Competition (FRC), a league involving three-versus-three matches, as dictated by a unique game released annually. The FIRST entry on \citet{wiki:first} provides a more detailed description of the organization and its history. 
Winning an individual match in the FRC requires one set of three robots, an alliance, scoring more points than an opposing alliance. Since its founding by Dean Kamen and Woodie Flowers in 1989, an increasing number of teams have designed robots according to each year's specific FRC game mechanics. The FRC has since ballooned to a pool of close to 4000 teams in 2018 playing in local and regional tournaments for a chance to qualify for the world championships of more than 800 teams. In turn, starting in 2017, FIRST divided its world championships into two championship tournaments in Houston and Detroit with about 400 robots playing in each.

 Important to this article's terminology and used frequently among scouters and tournament staff alike are the competing ``red'' and ``blue'' alliances to refer to the aforementioned sets of three robots. Those interpreting raw footage of matches rely heavily on the correspondingly colored bumpers of the robots in their data collection. Additionally, a robotics ``team'' refers to a group of people who have constructed, managed, and maintained a ``robot'', which refers to the actual machine playing in a match.
 
With the never-changing three-versus-three format arose questions posed by many FRC strategists over the years in analyzing past tournaments and their alliance selection phases: Which robots carried their alliance, and which robots were carried by their alliance? How might one use the answers to the first questions to predict hypothetical or upcoming matches?
 
\begin{figure}
 \centering
 \includegraphics[scale=0.65]{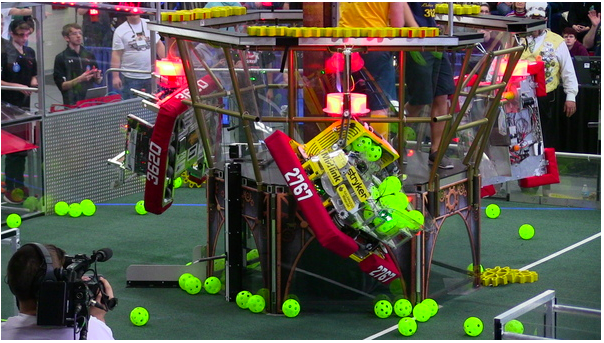}
 \caption{{\footnotesize Three robots on the red alliance demonstrate their ability to climb a rope and hold wiffle balls. The yellow gears were delivered from an angled slot or from the ground by the robots to be used in turning a crank on the top ledge. (Source: https://www.thebluealliance.com/team/2767/2017.) } }
 \label{fig:my_label}
\end{figure}

Over the years, FRC games have gone through a variety of changes and tasks that form a key part of competition. 
For 2017, tasks that earned points during a match included having the robot move across a line during the autonomous period, delivering game elements from one location to another using the robot, and having the robot climb a rope (Fig. \ref{fig:my_label}). Generally, tasks will be similar to these regardless of the year, though the game elements, delivery locations, and terrain will often change each year. {In addition to similar tasks}, there are some fundamental aspects that have not changed from year to year. Games across the years have followed the same match timing format of 15 seconds of autonomous control, in which robots are pre-programmed to operate in ways that score points, followed by a period of remote control, also called the teleoperated period, lasting at least two minutes. \citet{youtube:first} has provided a more detailed video guide of how points were scored for the 2017 FRC game, and also provides a similar video guide at the start of each season.

There also exist numerous regulations and loopholes to those regulations that affect the relationships of results between different tournaments. Before 2020, ideally, the same robot played in different tournaments would be expected to have the same scoring ability across those tournaments. However, analysts often consider robots of the same team from different tournaments independent from each other, since although FIRST implemented a ``Stop Build Day'' in which teams could no longer make significant changes to their robots, there were still loopholes that allowed for {a team's robot} to significantly improve in between tournaments. One such loophole allowed wealthier teams to build a secondary robot for testing new components, which could be added on during pre-tournament maintenance. As of 2020, the Stop Build Day rules have been lifted so that all teams can work on improvements more equitably in between tournaments. This development means that in future years, one can predict the performance difference of a robot between tournaments to be even more drastic than in recent history, thus reducing the reliability of previous tournament data in predicting the performance of a robot in future tournaments. Because of these developments and the history of inconsistency between tournaments, for the purposes of modeling, the same {team's entry} played in different tournaments will be considered a completely different robot. The model presented in this article is, thus, limited to the data from the qualification rounds to predict the {outcomes of the playoff rounds from the same tournament}. It must also be noted that most maintenance that occurs in between rounds does not have a significant additional positive impact on the robot's performance, since most of the scoring ability is inherently designed outside of competitive play.

The FRC's rules provide a structure with many features analogous to traditional sports. A kickoff event in January announces the rules of that year's game and signifies the beginning of the ``build season''. Teams often organize scrimmages or collaboratively test prototypes during this time in a fashion similar to roster formation during Spring Training in Major League Baseball (MLB), other preseasons, or even free agent workouts. Build season is followed by the competition season where good tournament results count toward advancement to increasingly higher levels of gameplay, most notably toward the aforementioned championship tournaments.

As of the 2018 competitions, each championship site is composed of six divisions. Each division runs a mini-tournament to determine a division champion alliance. The six division champion alliances further advance to the Einstein Division, a round-robin group whose top two alliances play a best-of-three to decide a site champion alliance. In some years, each site champion alliance plays the other at another gathering called the ``Festival of Champions''. This scheme somewhat parallels the MLB's National and American Leagues before the introduction of interleague play in that championship sites, championship divisions, and regions for qualifying for the world championships segregated teams from all over the world into their own pools of interaction, with the Festival of Champions serving as a {analog} to the MLB's World Series.

{As in all tournaments throughout a season, matches in each division mini-tournament are divided into a qualification stage and a playoff stage, which are similar to a microcosm of a regular season and a postseason in major sports leagues on the scale of only two days and with far more teams.} In the qualification stage, teams' robots are assigned by a predetermined algorithm into matches of six split into three in a blue alliance against another three in a red alliance. \citet{deshpande:2016} found similar types of data in the National Basketball Association (NBA), where the contributions of players are assessed through home team scores and away team scores at different shifts, which are defined to be periods of play between substitutions when ten players on the court is unchanged, during a match. Such a concept is also applicable in the National Hockey League (NHL).

Within the context of an individual game, the FRC also shares many similarities with traditional sports. Basketball is a game that could be broken into separate units such as possessions, in which one team has control of the ball and attempts to score by putting the ball in a basket, while the other team defends said basket. The above description defines a possession with a scoring method, a defense, and an offense. The FRC's {analog to the NBA's possession} is a cycling feature in which a robot may use the same method or set of actions to score repeatedly over the course of a match. Like the structure of a basketball possession, each of these cycles involves an offensive unit attempting to score with a specific action, and, sometimes, a defensive unit, which seeks to deter or defend from opponent scoring attempts. Robots on the same alliance may work in tandem within each cycle or in parallel, often running multiple independent cycles with the hopes of maximizing scoring. In both cases, a successful score for the offense is seen positively for offensive players and negatively for defensive players. That the FRC allows the possibility of simultaneous, independent cycles should not lessen the similarity between a cycle and a possession, since {each cycle} may be modeled the same way. The relation between cycles in the FRC and possessions in the NBA allows individual contributions of robots in a FIRST match to be analyzed like the individual contributions of players. {However, unlike the FRC, players in the NBA are not randomly assigned to shifts, and if a shift in the NBA is comparable to an individual match in the FRC, for any given NBA shift within a game, players of different teams cannot be teammates in traditional sports, whereas in the FRC, former teammates often end up facing each other in later matches on the same day.}

Especially worth mentioning is that for the FRC, since matches and playing field setup are standard without significant opportunity for fan interaction, there is no need to take into account a home-court advantage effect. A defensive component is also unnecessary in our model formulation. Since the designed robots are memory-less, scores of different matches are reasonably assumed to be mutually independent. 
There also exists the case in traditional sports where certain players may have so much synergy that a team simply does not provide the opportunity, and thus the data, for the hypothetical shifts where said players are separated. For the qualification stage of an FRC tournament, a predetermined schedule algorithm, as detailed in the ``match assignment" section of the 2017 game manual \citet{first:2017} provides more opportunities for different synergies to be tested and for the separation of two or three well partnered robots in the available data. Effectively, the FRC tournament design eliminates the analogous power of the coaches of traditional sports to influence the model's training data in less than ideal ways. It should be noted, however, that even though minute  differences exist in the updated wording of the algorithm between 2017's game and games after 2017, the algorithm itself has not changed. Although the scheduling algorithm is not truly random, it does eliminate many potential schedules that may be obviously unbalanced. Thus, the way the qualification round schedules are assigned only helps models account for irregularities in rating robots. {More importantly,} included in the eliminated potential schedules are those that repeat match combinations. {This control diversifies the head-to-head matchups between potential alliances, but since the scheduling algorithm forces uniqueness in matchmaking within a tournament and the same team's robot is considered a completely different entity between tournaments, every single qualification match across tournaments in an entire season is unique and unrepeatable. This makes simulating new matches very prone to error and hard to control, since any simulation of six robots in the same alliances would rely on only one pair of a red and blue score, if it exists.}

{It has been observed that within an alliance for a specific match, there is at least some robot on the winning alliance that contributed, possibly more than the others, to the win and at least some robot on the losing alliance that did not pull its weight, possibly more than the others, in the loss. For example, Team 254 of the 2018 Hopper Division went undefeated in its qualification matches, and Team 4775 of the same division failed to win a match in qualification 
per The Blue Alliance's online database \citet{tba:2018}. These, together, provides a motivation to estimate robot strengths to determine which robots are contributing the most to wins and losses. We aim in this article to develop a model that utilizes the estimated strengths of individual robots to predict the outcome of any match.}

Official ranking of FRC teams within the qualification stage involves a system of ranking points (RP). The kickoff event's revealing of game rules introduces two in-game objectives, the completion of which is rewarded with one RP each. Additionally, winning a match will net a team two {RPs} while losing produces no {additional} RP. Pre-playoff rankings are determined according to the total RP earned in the qualification stage with the top-8 teams in terms of RP being guaranteed a spot in the playoffs. These top-8 teams are given the ability to form playoff alliances with an additional three members for the playoff stage, one of which serves as a substitute, preserving the three-versus-three game structure. When there are ties within the RP system, the average score (AS) measure is used to break RP ties. 

Unlike the qualification stage, playoff stage alliances are determined by teams during the alliance selection phase. {Here,} the pool of teams within a particular tournament draft themselves into playoff alliances that they each believe have the best chances of winning in the playoff stage. It is after this phase that the new, team-selected alliances take full control of their own results, independent of randomness in having carried or being carried in the qualification stage.
The alliance selection phase, which takes place immediately after the end of qualification and directly affecting the upcoming playoff, also marks the transition between when RP is king (qualification) and when in-game scores become the ultimate goal (playoff). These reasons are why alliance selection is the most game-changing and consequential portion of an entire tournament.  Subsequently, there are some teams that design their robots specifically to complete the in-game objectives, important to obtaining a higher RP ranking, with the hopes of guaranteeing a spot in the playoffs and having more control over alliance selection. Teams that choose to implement this design strategy sometimes sacrifice scoring ability, which is more important during the playoff stage. However, sometimes this tradeoff for alliance selection control is understandable, since such teams usually invest more than others into scouting and analysis of the competition and would thus have an information advantage in making alliance selections. 

The alliance selection period is like the free agency period of the NBA. This is when players and managements, aside from in-season trades, negotiate the bulk of partnerships in, for non-rebuilding teams, hopes of capturing the best odds for the following season. It is important to note that unlike traditional sports leagues, the FRC does not have {an analog} to team-rebuilds within a tournament, so all alliance selections must be done with only the end goal of a championship in mind. Similarly, robotics teams, will negotiate with each other during the alliance selection phase to form the best alliance of robots for the playoff rounds. 
Synergy comes into play within both the free agency period and alliance selection, as management of an NBA team would not be looking to hypothetically field a team of five centers, nor would a FIRST team necessarily look to form a partnership with a team whose robot is optimized for the same scoring method. The best alliances tend to have each robot perform a different job, so as to maximize scoring opportunity, or minimize the other alliance's scoring opportunity if one of the jobs is general defense, depending on that year's game. Thus, NBA management and FRC strategists are comparable in that they both seek well-roundedness, as well as saturated ability in their quests for championships. Other traditional sports' free agencies also exhibit the same desires and goals from teams.

In the playoff stage, the eight alliances play three of their four robots in each match in a best of three matches knockout manner until a champion is declared. As the FRC continues to progress, so have questions from participants on improving the quality of alliance selection and decision making. Since the goal of playoff matches is actual points rather than RP, many decide not to rely on the RP system to make decisions regarding the alliance selection. Teams throughout the years have created new measures in order to assess actual or relative robot strengths. Some widely used measures in robotics forums \citep[e.g.][] {weingart:2006, law:2008,  gardner:2015, tba:opr} include the offensive power rating (OPR), the winning margin power rating (WMPR), and the calculated contribution to winning margin (CCWM), among others.  
In one such investigation of these measures, using simulated and past FRC and FTC tournaments, \citet{gardner:2015} further analyzed the performance of these measures.
However, the OPR and WMPR models were found to have poor predictive performance due to their imprecise estimates 
caused by a small ratio of the number of observations to the number of robots in the considered data. Furthermore, the CCWM model was shown to be a special case of the WMPR model and indicated to be meaningless in our application.

In the 2018 FRC Houston and Detroit championships, there were 112 to 114 matches involving 67 to 68 robots for divisions in the qualification stage. The corresponding ratios of the number of observations to the number of robots range from 3.28 to 3.40 for the OPR model and  from 1.64 to 1.70 for the WMPR model. The analysis of such paired comparison data also reveals no strong evidence to support the use of the WMPR model.
These considerations motivate us to explore some possible avenues to enhance their predictive capacities.
With the consideration of latent clusters of robot strengths, the proposed model can be regarded as the dimension reduction of the parameter spaces in the OPR and WMPR models. One crucial issue is how to estimate the number of clusters, clusters of robots, and robot strengths.The major aim of our proposal is to assess robot strengths in a more accurate and precise manner and help highly ranked teams assess potential alliance members for the opportunity of partnering in a playoff stage. To the best of our knowledge, there is still no research devoted to studying the latent clusters of individual strengths in team games.

The rest of this article is organized as follows. Section \ref{sec:lit-review} outlines existing measures and models for robot strengths. In Section \ref{sec:methodology}, we propose a more general regression model with latent clusters of robot strengths, develop two effective estimation procedures, and present some agreement and stability measures.
In Section \ref{sec:num-example}, our proposal is applied to the 2018 FRC Houston and Detroit championships. Conclusion and discussion are also given in Section \ref{sec:conclusion}.

\section{Existing Measures and Models for Robot Strengths} \label{sec:lit-review}

Throughout history, assessments for individual strengths have always intrigued analysts of team sports, e.g., individual stats of NBA players in basketball.
This is no different in the FRC as in the last 25 years, as the FRC game designs contain features and components analogous to those in other traditional sports.
Both FIRST and FRC teams have established their own systems for rating competing robots.
In this section, we concisely introduce and compare the OPR, CCWM, and WMPR measures and models
for robot strengths. Different from the OPR, CCWM, and WMPR models for the match outcome in the literature, a more general semi-parametric formulation is further presented in this study. Moreover, we review similar measures and models for individual strengths in other team games.

\subsection{Data and Notations}

Let $K$ and $M$ stand for the number of robots and the number of matches, respectively, in a division.
In each match $s$, three robots forming the blue alliance, $\text{B}_s$, go against three other robots forming the red alliance, $\text{R}_s$.
In the qualification stage, the first $\lceil \frac{1 \times K}{6} \rceil$ matches are designed to ensure that each robot plays {at least} one match, the succeeding 
$\lceil \frac{1 \times K}{6} \rceil$ {matches} are designed to ensure that each robot plays {at least} two matches, and in total, $M =\lceil \frac{m_0 \times K}{6} \rceil$ matches are designed to ensure that each robot plays {at least}  $m_0$ matches, where $\lceil$ $\rceil$ is the ceiling function.
For example, in the Carver division of the 2018 Houston championship, $K = 68$, $M =114$, and $m_0 = 10$.
With a total of 114 matches, each robot played at least ten matches. Since exactly four robots played eleven matches, per the rules as specified in the game manual \citet{first:2017}, the third match of each robot that played eleven games was not considered in their rankings, so that all robots have exactly ten matches worth of opportunity to perform.

To simplify the presentation, let $Y^{\text{B}}_s$ and $Y^{\text{R}}_s$ stand for the scores of $\text{B}_s$ and $\text{R}_s$, respectively; $\beta_{1}, \dots , \beta_{K}$ the strengths (or the relative strengths with the constraint $\sum^{K}_{i=1}\beta_{i} = 0$) of robots with the corresponding IDs $1,\dots,K$; and $i.i.d.$ the abbreviation of independent and identically distributed. The super-index in $\beta_{i}$'s  and their estimators is further used to distinguish different measures. For the linear model formulation of existing models, we also define the following notations:
\begin{eqnarray*}
 D_s &=& I\big(Y^{\text{R}}_s - Y^{\text{B}}_s > 0 \big),
 X^{\text{B}}_{si} = I(i \in \text{B}_s), X^{\text{R}}_{si} = I(i \in \text{R}_s),\hspace{2.5cm}\\
 Y^{(t)} &=& \big( Y^{(t)}_{1} , \dots ,  Y^{(t)}_M \big)^{\top}, \text{ and }
X^{(t)} = \big(X^{(t)}_{si}\big) = \big( X^{(t)}_{1} , \dots , X^{(t)}_M \big)^{\top} = \big( X^{(t)}_{(1)} , \dots , X^{(t)}_{(K)} \big),
\end{eqnarray*}
where $I(\cdot)$ is the indicator function, $X^{(t)}_{s}=(X^{(t)}_{s1},\dots, X^{(t)}_{sK})^{\top}$ and $X^{(t)}_{(i)}=(X^{(t)}_{1i},\dots,X^{(t)}_{Mi})^{\top}$ represent the covariate information of match $s$ and robot $i$, respectively, $s=1,\dots, M$, $i=1,\dots,K$, and $(t)$ can be either $\text{B}$ or $\text{R}$.

\subsection{Conceptual Models}\label{sec:CM}

In application, the AS, the average score, is a simple way to assess robot strengths. The AS of a robot is calculated by adding up the alliance scores in the matches played and dividing by three times of the number of matches played. Under this system,
the strength of robot $i$ is estimated by
\begin{eqnarray}
  \hat{\beta}_i= \frac{\sum_{s=1}^M \big(X^{\text{B}}_{si} Y^{\text{B}}_s +X^{\text{R}}_{si} Y^{\text{R}}_s \big)}{3\sum_{s=1}^M \big(X^{\text{B}}_{si} +X^{\text{R}}_{si}  \big)}, i = 1, \dots, K. \label{eqn:as}
\end{eqnarray}
Different from the plus-minus statistic, which was first adopted by the NHL's Montreal Canadiens in the 1950s, 
the AS is analogous to goals for or points for, instead of score differential, divided by the number of players.
Same as the drawback of the plus-minus statistic, obviously strong (weak) robots may be underrated (overrated) due to being paired with relatively weak (strong) robots, albeit robots are randomly assigned to matches.
Even with a large enough number of matches, the AS is still not a good representation for robot strengths.

As detailed by \citet{tba:opr}, in 2004, Karthik Kanagasabapthy of the FRC Team 1114 created a measure, which was termed the calculated contribution, to assess the contribution of a robot to an alliance score. He initiated work on the OPR measure, which characterizes the alliance score on the sum of contribution strengths of the alliance's robots, and estimated robot strengths 
by the least squares solution of systems of equations. The details of the computation were further explained in a post by \citet{weingart:2006}, who first called the calculated contribution as the OPR. In the following OPR model, the blue alliance score and the red alliance score in each match are formulated in the same way:
\begin{eqnarray}
 Y^{\text{R}}_s = \sum_{\{i \in \text{R}_s\}} \beta_{i} + \varepsilon^{\text{R}}_s
  \text{ and }
  Y^{\text{B}}_s= \sum_{\{i \in \text{B}_s\}} \beta_{i} + \varepsilon^{\text{B}}_s,
 s = 1, \dots, M, \label{eqn:opr}
\end{eqnarray}
where  $\varepsilon^{\text{R}}_1 , \dots , \varepsilon^{\text{R}}_M, \varepsilon^{\text{B}}_1 , \dots , \varepsilon^{\text{B}}_M$ are $i.i.d.$ with mean zero and variance $\sigma^2$.
As we can see, the above formulation is similar to that from \citet{macdonald:2011} for NHL players except that the OPR model does not consider a home-court advantage effect  and a defensive component.
Under the assumption of independent errors,
$2M$ observations are used in the estimation of $\beta_{1}, \dots , \beta_{K}$.
While such a model formulation allows for the decomposition of expected alliance scores into individual robot strengths in a logical manner, it completely ignores actions of robots in the opposing alliance. As we can see, the independence between $\varepsilon^{\text{R}}_s$ and $\varepsilon^{\text{B}}_s$, $s=1,\dots, M$, is unrealistic in practice.
Furthermore, robots in blue and red alliances might be influenced by a common unobserved factor (e.g. environmental obstacles), say $Z_{s}$, in match $s$ such that $E[Z_{s}|\{i:i \in \text{B}_s \text{ or } \text{R}_s\}]=E[Z_{s}]=\beta_{0}\neq 0$, $s=1,\dots, M$. To achieve this more realistic interpretation, the OPR model can be modified as
\begin{eqnarray}
Y^{\text{R}}_s = \beta_{0}+\sum_{\{i \in \text{R}_s\}} \beta_{i} + \varepsilon^{\text{R}}_s
 \text{ and }
 Y^{\text{B}}_s =\beta_{0}+ \sum_{\{i \in \text{B}_s\}} \beta_{i} + \varepsilon^{\text{B}}_s,
 s = 1, \dots, M, \label{eqn:opr1}
\end{eqnarray}
where the intercept $\beta_{0}$ can be interpreted as the average of all robot strengths. Furthermore, the error terms $\varepsilon^{\text{R}}_s$ and 
$\varepsilon^{\text{B}}_s$ cannot be assumed to be mutually independent, $s=1,\dots, M$.
It is notable that the resulting estimators of the regression coefficients in model (\ref{eqn:opr})
are biased estimators of those in model (\ref{eqn:opr1}).

Another commonly used assessment for robot strengths is the WMPR measure from \cite{gardner:2015}.
This measure accounts for the effects of robots on the opposing alliance.
With this consideration, the difference in scores between alliances in the WMPR model is formulated as
\begin{eqnarray}
 Y^{\text{R}}_s  -Y^{\text{B}}_s = \sum_{\{i \in \text{R}_s\}} \beta_{i} - \sum_{\{i \in \text{B}_s\}} \beta_{i} + \varepsilon_{s}, s = 1, \dots, M, \label{eqn:wmpr}
\end{eqnarray}
where $\varepsilon_{1} , \dots , \varepsilon_{M}$ are $i.i.d.$ with mean zero and variance $\sigma^2$.
In basketball, soccer, volleyball, and esport games, some research works such as \citet{rosenbaum:2004}, \citet{macdonald:2011}, \citet{schuckers:2011}, \citet{saebo:2015}, \citet{hass:2018}, and \citet{clark:2020}, among others, used the adjusted plus/minus rating (APMR), which is similar to the WMPR, to assess the contribution to net scores per possession by each player relative to all other players. In contrast with the WMPR model, the APMR model takes into account a home-court advantage and clutch time/garbage time play, typical of many team sports.
In traditional sports, this model formulation has been shown to be useful for paired comparison data.
A more general condition is imposed on $\varepsilon_{1}, \dots , \varepsilon_{M}$, albeit the OPR model in (\ref{eqn:opr}) or (\ref{eqn:opr1}) also leads to
\begin{eqnarray}
 Y^{\text{R}}_s  -Y^{\text{B}}_s = \sum_{\{i \in \text{R}_s\}} \beta_{i} - \sum_{\{i \in \text{B}_s\}} \beta_{i} + \big(\varepsilon^{\text{R}}_s - \varepsilon^{\text{B}}_s\big), s = 1, \dots, M, \label{eqn:wmpr1}
\end{eqnarray}
where $(\varepsilon^{\text{R}}_1 - \varepsilon^{\text{B}}_1 ), \dots , (\varepsilon^{\text{R}}_M - \varepsilon^{\text{B}}_M)$ 
are $i.i.d.$ with mean zero and variance $E[(\varepsilon^{\text{R}}_1 - \varepsilon^{\text{B}}_1 )^{2}]$.
However, only $M$ observations are used in the estimation of robot strengths.

By taking into account the influence of the opposing alliance score, \citet{law:2008} showed a CCWM model of the following form:
\begin{eqnarray}
Y^{\text{R}}_s  -Y^{\text{B}}_s = \sum_{\{i \in \text{R}_s \}} \beta_{i} + \varepsilon^{\text{R}}_s
\text{ and }
Y^{\text{B}}_s  -Y^{\text{R}}_s  = \sum_{\{i \in \text{B}_s\}} \beta_{i} + \varepsilon^{\text{B}}_s, s = 1, \dots, M, \label{eqn:ccwm}
\end{eqnarray}
where $\varepsilon^{\text{R}}_1 , \dots , \varepsilon^{\text{R}}_M, \varepsilon^{\text{B}}_1 , \dots , \varepsilon^{\text{B}}_M$  
are $i.i.d.$ with mean zero and variance $\sigma^2$.
Clearly, the CCWM model formulates the net effect of the opposing alliance score on the reference alliance score.
The contribution of a robot in its participated match is further explained by the winning margin.
However, by the equality $E[Y^{\text{R}}_s  -Y^{\text{B}}_s] = - E[Y^{\text{B}}_s  -Y^{\text{R}}_s ]$, we have
$\sum_{\{ i \in \text{R}_s\}} \beta_{i} = -\sum_{\{ i \in \text{B}_s\}} \beta_{i}$, $s=1,\dots, M$, which leads to $\beta_{1} = \dots = \beta_{K} = 0$ 
for $M\geq K$ and $ \varepsilon^{\text{R}}_s=- \varepsilon^{\text{B}}_s, s=1,\dots, M$.
As a result,
\begin{eqnarray}
Y^{\text{R}}_s  -Y^{\text{B}}_s =  \varepsilon_{s}, s = 1, \dots, M, \label{eqn:ccwm1}
\end{eqnarray}
where $\varepsilon_{1},\dots,\varepsilon_{M}$ are $i.i.d.$ with mean zero and variance $\sigma^2$.
This fact indicates that the CCWM model is meaningless in our application.
Obviously, the model formulation of the CCWM model in (\ref{eqn:ccwm1})
is a special case of that in (\ref{eqn:wmpr}) with $\beta_{1}=\dots=\beta_{K}=0$. 
For this reason, the CCWM model will not be studied in the rest of this article.

\begin{rem}
For the defensive strengths of robots, there were some proposals to take into account the defensive power rating (DPR). It was shown by \citet{gardner:2015} that estimated robot strengths of the DPR model can be expressed as the difference of those of the OPR and CCWM models. The author further introduced other measures such as combined power rating, mixture-based ether power rating, and some related simultaneous measures. However, in the setup of the FRC system, these measures are inappropriate to characterize robot strengths. For this reason, we do not explore their properties and extensions in this article.
\qed
\end{rem}

\subsection{Comparisons to Measures in Other Team Games}

Wins above replacement (WAR), as described by \citet{slowinski:2010}, and value over replacement player (VORP), detailed by \citet{woolner:2001a, woolner:2001b},  are statistics used in baseball to express individual contributions of a player to his team. WAR measures this by wins in a 162-game season, while VORP measures offensive contributions of a player through runs created and pitching contributions through allowed run averages. These statistics are comparable to the different individual contribution statistics analyzed in this article. Out of these two statistics, VORP is more similar to the currently available robotics statistics, since the lower number of matches in qualification, which does not usually pass 11 per robot for the 2018 championships, means there is not enough data to fit an accurate WAR-like measure.

While the APMR model seems to gain an advantage in describing the basketball, ice hockey, soccer, volleyball, and esport games, the context is quite different from robotics competitions.
Furthermore, \citet{ilardi:2008} incorporated the role of each player as offensive or defensive into the APMR model. Based on the difference between the expected and observed outcomes, \citet{schuckers:2011} proposed an adjusted plus-minus probability model.
In a robotics tournament, the ratio of the number of observations to the number of robots in each division is under two for the WMPR model, while the ratio of the number of possessions to the number of NBA players in a season is about sixteen (see \citet{teamrankings:2019} for the 2018-2019 season) for the APMR model. This limitation in the WMPR model usually produces poor estimation of robot strengths and poor prediction of future outcomes. An important task of our research is, thus, to develop a better predictive model.

\subsection{Linear Model Formulation} \label{sec:RMF}

It follows from (\ref{eqn:opr}) and (\ref{eqn:wmpr}) that the OPR and WMPR models can be expressed as
\begin{eqnarray}
Y = X \beta+ \varepsilon, \label{eqn:linear}
\end{eqnarray}
where $\beta = \big( \beta_{1} , \dots , \beta_{K} \big)^T$
and $\varepsilon$ has mean vector $0_{\bar{M} \times 1}$ and covariance matrix $\sigma^2 I_{\bar{M}}$. In terms of such a linear model formulation, we have
\begin{enumerate}
\item $Y =\begin{pmatrix}Y^{\text{R}} \\ Y^{\text{B}} \end{pmatrix}$, $X =\begin{pmatrix} X^{\text{R}} \\ X^{\text{B}} \end{pmatrix}$, and $\varepsilon =\big(\varepsilon^{\text{R}}_1, \dots , \varepsilon^{\text{R}}_M,\varepsilon^{\text{B}}_1, \dots ,\varepsilon^{\text{B}}_M\big)^\top$ with $\bar{M} = 2M$ in the OPR model; and
\item $Y = Y^{\text{R}} - Y^{\text{B}}$, $X = X^{\text{R}} - X^{\text{B}}$, and $\varepsilon = (\varepsilon_1, \dots ,$ $\varepsilon_M)^T$ with $\bar{M} = M$ in the WMPR model.
\end{enumerate}
Owing to the linear dependence of $\big(X^{\text{R}}_{(i)}-X^{\text{B}}_{(i)}\big)$'s (i.e. $\sum^{K}_{i=1}\big(X^{\text{R}}_{(i)}-X^{\text{B}}_{(i)}\big)=0$) in the WMPR model, the constraint $\sum^K_{i=1} \beta_{i} = 0$, which was also adopted by \citet{cattelan:2013} for the Bradley-Terry specification \citep{bradley:1952, bradley:1953}, is further imposed to solve the identifiability of $\beta$. As a result, the coefficient $\beta_{i}$, compared to $\beta_{K}$, is explained as the relative strength of robot $i$, $i=1,\dots, K-1$.
With this constraint, model (\ref{eqn:linear}) can be rewritten as
\begin{eqnarray}
Y =\bar{X} \beta^{*}+ \varepsilon,
\label{eqn:linear.wmpr}
\end{eqnarray}
where $\bar{X}$ is a $M\times (K-1)$ matrix with the $i$th column being $X_{(i)}-X_{(K)}$ and $\beta^{*}=\big(\beta_{1},\dots,\beta_{K-1}\big)^{\top}$ with $\beta_{K}=-\sum^{K-1}_{i=1} \beta_{i}$. In the context of regression analysis, $\beta$ (or $\beta^{*}$) is naturally estimated by the least squares estimator (LSE), say $\widehat{\beta}$ (or $\widehat{\beta}^{*}$), under the Gauss-Markov conditions.

\begin{rem}
Under the WMPR model, \citet{gardner:2015} proposed an estimator of $\beta$ by solving the pseudo-inverse solution of the corresponding estimating equations of the sum of squares. Albeit lacking an explanation for the resulting estimator, the estimated or predicted score of a match is unique.
In application,
the constraint $\sum^{K}_{i=1}\beta^{2}_{i}=1$ can also be imposed to obtain estimators of relative robot strengths. However, there is no simple formulation for the corresponding LSE of $\beta$. \qed
\end{rem}

\begin{rem}
Like the  development of a plus/minus rating system by \citet{feamhead:2011}, the robot strengths $\beta_{i}$'s in the WMPR model can also be formulated as $i.i.d.$ 
normal random variables with mean zero and variance $\sigma^{2}$. Together with the normality assumption on
$\varepsilon$ and the independence between $\beta$ and $\varepsilon$, the resulting predictor of $\beta$ 
has been shown by \citet{fahrmeir:2001} to be a ridge estimator, which is the same as that in \citet{sill:2010}, with the regularization parameter $\sigma^{2}/\sigma^{2}_{0}$ in such a Bayesian framework. Since the ridge regression is mainly used to combat multicollinearity of covariates, its explanation is different from that of the introduced Bayes estimator. Furthermore, in most existing paired comparisons models, this problem was solved by imposing the constraint  
$\sum^{K}_{i=1}\beta_{i}=0$. In our application to the 2018 FRC championships, it is also shown that the WMPR model and its Bayesian formulation, which is hereinafter referred to as the WMPRR model, have comparable performance in prediction.
\qed
\end{rem}

\subsection{Binary Regression Models}  \label{sec:BRMF}

It is implied by the OPR and WMPR models that the corresponding conditional probabilities of a match outcome have the following form:
\begin{eqnarray}
P\big(D_s = 1 | X^{\text{B}}_{s} = x^{\text{B}}_{s} , X^{\text{R}}_{s} = x^{\text{R}}_{s}\big) = 1 - F \big( - (x^{\text{R}}_s - x^{\text{B}}_s)^\top \beta \big), s=1,\dots, M,\label{eqn:bin}
\end{eqnarray}
where $F(\cdot)$ is an unknown distribution function. For the OPR model, $F(\cdot)$ is a distribution function of $(\varepsilon^{\text{R}}_s - \varepsilon^{\text{B}}_s)$'s. Although the conditional probability in (\ref{eqn:bin}) can also be derived from model (\ref{eqn:opr1}),
the LSEs of the regression coefficients in both models and the resulting nonparametric estimators of $F(\cdot)$
will be completely different. As for the WMPR model, $F(\cdot)$ is a distribution function of $\varepsilon_i$'s.

Let  $(Y^{\text{B}}_{0},Y^{\text{R}}_{0},X^{\text{B}}_{0},X^{\text{R}}_{0})$ stand for a future run
with $D_{0}=I\big(Y^{\text{R}}_{0}-Y^{\text{B}}_{0}>0\big)$ and $(x^{\text{B}}_{0},x^{\text{R}}_{0})$ being a realization of $(X^{\text{B}}_{0},X^{\text{R}}_{0})$.
In practical implementation, we will conclude that
\begin{eqnarray}
\{ Y^{\text{R}}_0 > Y^{\text{B}}_0\} \text{ if } F \big( - (x^{\text{R}}_0 - x^{\text{B}}_0)^\top \beta \big)<0.5 \text{ and }\{ Y^{\text{R}}_0 \leq Y^{\text{B}}_0\} \text{ otherwise.} \label{eqn:discrim}
\end{eqnarray}
Especially worth mentioning is that {we do not assume any structure} on $F(\cdot)$. For a strict monotonicity of $F(u)$ with $F(0) = 0.5$, $F \big( - (x^{\text{R}}_0 - x^{\text{B}}_0)^\top \beta \big)<0.5$ is equivalent to
$\big( x^{\text{R}}_0 - x^{\text{B}}_0 \big)^\top \beta> 0$. Special cases of $F(u)$ for paired comparison data include the logistic distribution function
and the standard normal distribution function, which lead to the Bradley-Terry model \citep{bradley:1952,  bradley:1953} and the Thurstone-Mosteller model  \citep{thurstone:1927, mosteller:1951}.
In various sports and games, the Bradley-Terry model has been widely used to rate players.
It includes the studies of \citet{cattelan:2013} in basketball and football, \citet{elo:1978} in chess and other sports, \citet{islam:2017} in cricket, \citet{chen:2017} and \citet{weng:2011} in online multi-player games, and \citet{huang:2008} in bridge. With the specifications of the complementary logistic function for the ``update function shape" and the logistic model for the margin of victory in the Elo-type rating systems, \citet{aldous:2017} and \citet{kovalchik:2020}, respectively, showed that the resulting models are related to the Bradley-Terry model.

\section{Latent Clusters of Robot Strengths} \label{sec:methodology}

A more general model formulation, compared to existing models, is used to characterize latent clusters of robot strengths.
Two effective estimation procedures are further developed for the proposed regression model. Moreover,
some measures for the predictive ability of models, the agreement related to FRC ratings and model-based robot strengths, and the stability of estimated robot strengths and accuracies are presented in this section.

\subsection{Model Extensions}

\begin{figure}[htbp]
\centering
 \includegraphics[width=7cm,height=4.5cm]{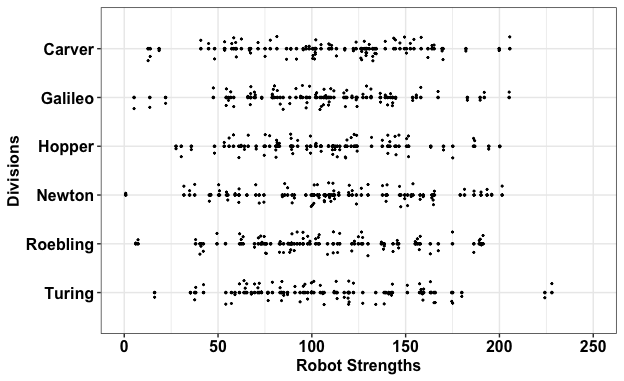}
 \includegraphics[width=7cm,height=4.5cm]{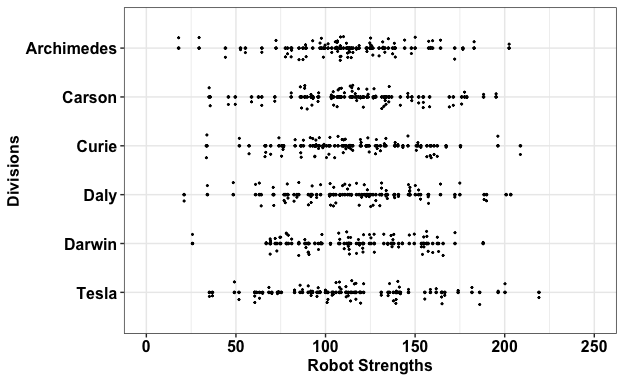}
  \caption{{\footnotesize Dot plots of estimated robot strengths from the OPR model for the divisions of the 2018 FRC Houston and Detroit championships with vertical jitter.}} \label{Cen-HouDet-OPR}
\end{figure}

\begin{figure}[htbp]
\centering
 \includegraphics[width=7cm,height=4.5cm]{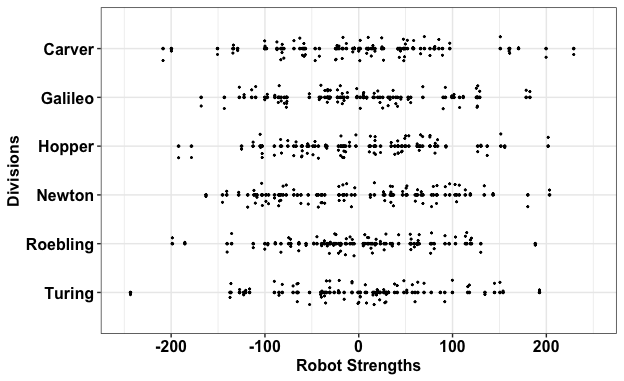}
  \includegraphics[width=7cm,height=4.5cm]{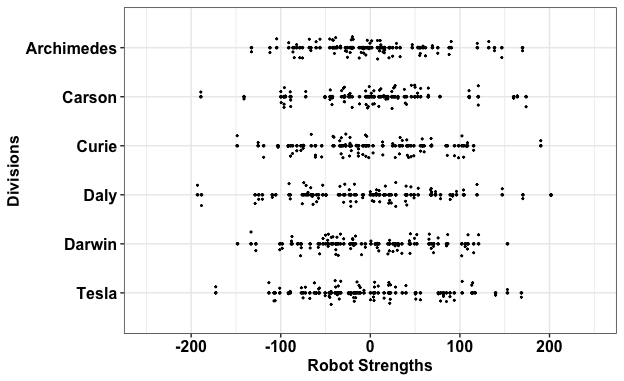}
 \caption{{\footnotesize Dot plots of estimated robot strengths from the WMPR model for the divisions of the 2018 FRC Houston and Detroit championships with vertical jitter.}}\label{Cen-HouDet-WMPR}
  \end{figure}

{In Fig.  \ref{Cen-HouDet-OPR} and Fig. \ref{Cen-HouDet-WMPR}, we observe a clustering feature in
the estimated robot strengths from the OPR and WMPR models for these two championship tournaments. With this finding, both models are extended to a more general formulation with a  clustering feature of robot strengths,
i.e.
\[
\beta = \Big(\beta^{c_0}_{g_1}, \dots , \beta^{c_0}_{g_K}\Big)^{\top},
\]
where $c_0$ is an unknown number of clusters and $g_i
\in \{1, \dots , c_0 \}$  is the corresponding cluster of robot $i$, $i = 1, \dots , K$, with $g=(g_{1},\dots,g_{K})^{\top}$. 
In addition to reducing the number of parameters in the OPR and WMPR models, we do not assume any particular distribution for the underlying distributions of 
the difference in scores and the match outcome (win/loss). In fact, the viewpoint of  grouping  players into some skill levels has been adopted by the United States Chess Federation (USCF) and the National Wheelchair Basketball Association (NWBA). By using a rating system as defined by \cite{elo:1978},
which is used to assess the relative skill levels of players, the USCF classifies chess players into thirteen groups: class A, $\dots$, class J, expert or candidate master, national master, and senior master. Based on the player's physical capacity to execute fundamental basketball movements, the NWBA groups players into one of eight categories (1, 1.5, $\dots$, 4, and 4.5). However, these classification criteria are slightly subjective in practical implementation.}

By incorporating the information of $\beta_{i}=\beta^{c_0}_{g_i}$ with $g_{i}\in\{1,\dots,c_{0}\}$, $i=1,\dots, K$,  into the model formulation $Y=\beta_{1}X_{(1)}+\dots+\beta_{K}X_{(K)}+\varepsilon$ in (\ref{eqn:linear}), the covariates $X_{(i)}$'s sharing the same coefficient, say $\beta^{c_0}_{j}$, can be further summed together to produce a new covariate $X^{c_{0}}_{(j)}$, $j=1,\dots,c_{0}$. It
follows that $Y$ can be expressed as $\beta^{c_0}_{1}X^{c_0}_{(1)}+\dots+\beta^{c_0}_{c_0}X^{c_0}_{(c_0)}+\varepsilon$ and model (\ref{eqn:linear}) is an overparameterized version of model (\ref{eqn:linearcl}) for $K>c_{0}$.

Based on theoretical and practical considerations, an extended linear regression model is, thus, proposed as follows:
\begin{align}
Y = X^{c_0} \beta^{c_0}+ \varepsilon, \label{eqn:linearcl}
\end{align}
where $\beta^{c_0}= ( \beta^{c_0}_{1}, \dots , \beta^{c_0}_{c_0})^\top$, $\varepsilon$ has mean vector $ 0_{\bar{M} \times 1}$ and covariance matrix $\sigma^2_0 I_{\bar{M}}$, and
$X^{c_0}=( X^{c_0}_{1}, \dots , X^{c_0}_{M} )^\top= ( X^{c_0}_{(1)}, \dots , X^{c_0}_{(c_0)} )$ is a designed covariate matrix of
$X^{Bc_0}$ and $X^{Rc_0}$ with $X^{\text{B}c_{0}}_{(j)}$ and $X^{\text{R}c_{0}}_{(j)}$ being defined as $X^{c_{0}}_{(j)}$ based on $X^{\text{B}}_{(i)}$'s
and $X^{\text{R}}_{(i)}$'s, respectively.
Note that model (\ref{eqn:linear}) is our special case with $c_0 = K$ and is over-parameterized when $c_{0}$ is smaller than $K$.
In addition, to generalize the model formulation of robot strengths in (\ref{eqn:linear}), {we do not assume any particular distribution for the error term $\varepsilon$.} Since the ratio of the number of matches to the number of robots in each division is rather small, our study provides a possible avenue to enhance the predictive capacities of existing models.
Same with the derivation in (\ref{eqn:bin}), we have
\begin{eqnarray}
P\big(D_s = 1 | X^{\text{B}}_{s} = x^{\text{B}}_{s} , X^{\text{R}}_{s} = x^{\text{R}}_{s}\big)
= 1 - F\big( - (x^{\text{R}c_0}_s - x^{\text{B}c_0}_s)^T \beta^{c_0} \big), s=1,\dots, M,\hspace{0.03in}\label{eqn:Ebin}
\end{eqnarray}
for the OPR model with latent clusters of robot strengths (OPRC) and
the WMPR model with latent clusters of robot strengths (WMPRC).

\begin{rem}
Under the assumption that chess players of comparable skill levels play against each other, \citet{sismanis:2010} developed the Elo$++$ rating system to avoid overfitting the ratings of players in the Elo rating system. Since robots are randomly assigned to matches and alliances in the qualification stage, there is no explicit neighbor for each robot and, hence, this technique is inappropriate for estimating the OPRC and WMPRC models. Furthermore, it is rather subjective for the defined neighbor averages in the introduced $\ell_{2}$--regularization. 
\qed
\end{rem}

\subsection{Predictive Ability} \label{sec:APA}

In this subsection, we consider all possible regression models of the form
\begin{eqnarray}
Y = X^{c} \beta^{c} + \varepsilon, \label{eqn:linearcl1}
\end{eqnarray}
where $X^{c}=(X^{c}_{(1)},\dots, X^{c}_{(c)})$ and $\beta^{c}=(\beta^{c}_{1},\dots,\beta^{c}_{c})^{\top}$ with $X^{c}_{(j)}$ being defined as $X^{c_0}_{(j)}$
based on $X_{(i)}$'s, $c= 1, \dots , K$.
As in many scientific studies, sensitivity and specificity are commonly used measures of diagnostic accuracy. We further employ these two measures to detect the condition when the blue alliance defeats the red alliance and the condition when the red alliance defeats the blue alliance, respectively. Since robots are randomly assigned to one of two alliances, the expected value of the win indicator $D_{0}$ should be 0.5. It follows that the accuracy, i.e. the weighted  average of sensitivity and specificity, is a reasonable measure to assess the predictive ability of a model. More precisely, such a measure is the expected proportion of correct predictions.
Given the number of clusters $c$, clusters of robots $\widehat{g}$, and an estimator $\widehat{\beta}^{c}$ of $\beta^{c}$, the accuracy of a predictor based on model (\ref{eqn:linearcl1}) is defined as
\begin{eqnarray}
\text{AC}(c) &=& P\big( \text{sgn}\big(Y^{\text{R}}_0 - Y^{\text{B}}_0\big) \cdot \text{sgn}\big(\widehat{P}^{c}\big(x^{\text{B}}_0 , x^{\text{R}}_0\big) - 0.5\big) > 0 \big) \nonumber\\
&&+0.5 P\big( \text{sgn}\big(Y^{\text{R}}_0 - Y^{\text{B}}_0\big) \cdot \text{sgn}\big(\widehat{P}^{c}\big(x^{\text{B}}_0 , x^{\text{R}}_0\big) - 0.5\big) =0 \big), \label{eqn:PR}
\end{eqnarray}
{where $\text{sgn}(\cdot)$ is the signum function} and $\widehat{P}^{c}(x^{\text{B}}_0 , x^{\text{R}}_0) = 1 - \widehat{F}^{c}
\big( - (x^{\text{R}c}_0 - x^{\text{B}c}_0)^\top \widehat{\beta}^{c} \big)$ with
$\widehat{F}^{c}(v) = \sum_{s=1}^{M} I(e^{c}_s \leq v) /m$  and $e^{c}_s = (Y^{\text{R}}_s - Y^{\text{B}}_s) - (X^{\text{R}c}_s - X^{\text{B}c}_s)^\top \widehat{\beta}^{(c)}$,
$c = 1, \dots , K$, $s = 1 , \dots , M$. We note that the second term on the right-hand side of (\ref{eqn:PR}) is used to deal with the problem of ties.
Provided that $c_{0}$ and $g$ are known, the empirical distribution function $\widehat{F}^{c_0}(v)$ of residuals is shown by \citet{mammen:1996} to converge uniformly to the distribution function of errors under some suitable conditions.
For the difference in score between the red and blue alliances, i.e.  $(Y^{\text{R}}_0 - Y^{\text{B}}_0)$, the following mean square prediction error is adopted:
\begin{eqnarray}
\text{MSPE}(c) = E\big[ \big(\big(Y^{\text{R}}_0 - Y^{\text{B}}_0\big) - \big(x^{\text{R}c}_0 - x^{\text{B}c}_0\big)^\top \widehat{\beta}^{c} \big)^2 \big],~c = 1 , \dots , K. \label{eqn:MSPE}
\end{eqnarray}
 As in the context of regression analysis, this measure is found to be useful in exploring the effect of potential influential observations and
 in selecting several competing models.

In the qualification stage, $\text{AC}(c)$ and $\text{MSPE}(c)$ are estimated by the leave-one-match-out cross-validation estimates 
\begin{equation}
\widehat{\text{AC}}(c) = \frac{1}{M} \sum_{s=1}^{M} D_{s}(c)
\end{equation}
and
\begin{equation}
\widehat{\text{MSPE}}(c) = \frac{1}{M} \sum_{s=1}^{M}  \big((Y^{\text{\text{R}}}_s - Y^{\text{\text{B}}}_s) -  (X^{\text{R}c}_s - X^{\text{B}c}_s)^\top \widehat{\beta}^{c}_{-s} \big)^2, \text{ respectively, } c = 1 , \dots , K,
\label{eqn:PRcv}
\end{equation}
where 
\begin{eqnarray*}
D_{s}(c)&=&  I \big( \text{sgn}(Y^{\text{R}}_s - Y^{\text{B}}_s) \cdot \text{sgn}(\widehat{P}^{c}_{-s}(X^{\text{B}}_s , X^{\text{R}}_s) - 0.5) > 0\big) \nonumber \\
&&+ 0.5 I \big( \text{sgn}(Y^{\text{R}}_s - Y^{\text{B}}_s) \cdot \text{sgn}(\widehat{P}^{c}_{-s}(X^{\text{B}}_s , X^{\text{R}}_s) - 0.5) = 0\big), \\
\widehat{P}^{c}_{-s}(X^{\text{B}}_s , X^{\text{R}}_s) &=& 1 - \widehat{F}^{c}_{-s} \big( - (X^{\text{R}c}_s - X^{\text{B}c}_s)^\top \widehat{\beta}^{c}_{-s} \big), 
\end{eqnarray*}
and $\big(\widehat{\beta}^{c}_{-s},\widehat{F}^{c}_{-s}(v)\big)$ is computed as $\big(\widehat{\beta}^{c},\widehat{F}^{c}(v)\big)$ with match $s$ being removed, $s = 1, \dots , M$. Instead of using data
$\{Y_{-s},X^{c}_{-s}\}$, $\big(\widehat{\beta}^{c}_{-s},\widehat{F}^{c}_{-s}(v)\big)$ can be directly obtained through the relation between $\widehat{\beta}^{c}_{-s}$ and $\{\widehat{\beta}^{c},X^{c},e^{c}_{s}\}$ for each $s$. The computational details are relegated to Appendix A. {Let $\text{AC}_{1}$ and  $\text{AC}_{2}$ stand for the accuracies of any two generic models of a division with the corresponding estimates $\widehat{\text{AC}}_{1} = \sum_{s=1}^{M} D_{1s}/M$ and $\widehat{\text{AC}}_{2} = \sum_{s=1}^{M} D_{2s}/M$, where $D_{1s}$'s and $D_{2s}$'s are defined as $D_{s}(c)$'s. 
It is ensured by Theorem 1 in \cite{chiang:2012} that $M(\widehat{\text{AC}}_{\ell} -\text{AC}_{\ell})/\sqrt{\sum^{M}_{s=1} (D_{\ell s}-\widehat{\text{AC}}_{\ell})^{2}}$, $\ell=1,2,$ can be approximated by the standard normal {distribution}. When there is no tie in $D_{\ell s}$'s the standard error of $\widehat{\text{AC}}_{\ell}$ is further simplified to $\sqrt{\widehat{\text{AC}}(1-\widehat{\text{AC}})/M}$.
Using this property, an approximate $(1-\alpha)$-confidence interval for $\text{AC}_{\ell}$ is, thus, constructed as follows:
\begin{equation}
\widehat{\text{AC}}_{\ell}\pm \frac{ z_{1-\alpha/2}}{M}\sqrt{\sum^{M}_{s=1} \big(D_{\ell s}-\widehat{\text{AC}}_{\ell}\big)^{2}},~\ell=1,2,\label{eqn:ci}
\end{equation}
where where $z_{q}$ is the $q$th quantile value of the standard normal distribution.
For the hypotheses  $H_{0}: \text{AC}_{1}=\text{AC}_{2}$ versus $H_{A}: \text{AC}_{1}\neq \text{AC}_{2}$ (or the hypotheses $H_{0}: \text{AC}_{1}\leq\text{AC}_{2}$
versus $H_{A}: \text{AC}_{1}>\text{AC}_{2}$), the following test statistic is proposed:
\begin{equation}
T=\frac{M\big(\widehat{\text{AC}}_{1}-\widehat{\text{AC}}_{2}\big)}{\sqrt{\sum^{M}_{s=1} \big((D_{1s}-D_{2s})-(\widehat{\text{AC}}_{1}-\widehat{\text{AC}}_{2})\big)^{2}} }.\label{eqn:test}
\end{equation}
In our test, $H_{0}$ is rejected with an approximate  significance level $\alpha$ whenever $|T|\leq Z_{1-\alpha/2}$ (or $T>Z_{1-\alpha}$).}

{In the playoff stage, we let 
$K^{*}$ and $M^{*}$ stand for the number of robots and the number of matches, respectively, in a division; and 
$Y^{*(t)} = \big( Y^{*(t)}_{1} , \dots , Y^{*(t)}_{M^*} \big)^\top$ and 
$X^{*(t)} = \big( X^{*(t)}_{1} , \dots , X^{*(t)}_{M^*} \big)^\top$ the designed response vector and covariate matrix with
$(t)$ being either $\text{B}$ or $\text{R}$.}
 {By treating qualification data as training data, the measures $\text{AC}(c)$ and $\text{MSPE}(c)$ of an estimation procedure on playoff data are defined as} 
\begin{equation}
\text{AC}(c)=\frac{1}{M^*} \sum_{s=1}^{M^{*}} D^{*}_{s}(c) 
\end{equation}\text{ and }
\begin{equation}
\text{MSPE}(c) = \frac{1}{M^{*}} \sum_{s=1}^{M^{*}}  \big((Y^{*\text{R}}_{s} - Y^{*\text{B}}_{s}) -  (X^{*\text{R}c}_{s} - X^{*\text{B}c}_{s})^\top \widehat{\beta}^{c} \big)^2, c = 1 , \dots , K^{*}, \label{eqn:PRcvpo}
\end{equation}
where {$D^{*}_{s}(c)= I \big( \text{sgn}(Y^{*\text{R}}_{s} - Y^{*\text{B}}_{s}) \cdot \text{sgn}(\widehat{P}^{c}(X^{*\text{B}c}_{s} , X^{*\text{R}c}_{s}) - 0.5) > 0\big)
+0.5 I \big( \text{sgn}(Y^{*\text{R}}_{s} - Y^{*\text{B}}_{s}) \cdot \text{sgn}(\widehat{P}^{c}(X^{*\text{B}c}_{s} , X^{*\text{R}c}_{s}) - 0.5) = 0\big)$
with $(X^{*\text{B}c}_{s} , X^{*\text{R}c}_{s})$'s being computed as $(X^{\text{B}c}_{s} , X^{\text{R}c}_{s})$'s.}
As shown in our application, the predictive ability of the current and proposed models is poor for match outcomes in the playoff stage. Due to non-random assignment of robots to matches, some confounders cannot be treated merely as nuisance variables in model fitting. This partly explains poor capacity in prediction.

\subsection{Estimation Procedures}

For the unknown parameters $c_0$, $g$, and $\beta^{c_0}$ in model (\ref{eqn:linearcl}), it is usually impractical to fit all possible regression models in
(\ref{eqn:linearcl1}).
The total number of these model candidates can be further shown to be
\begin{eqnarray}
S_K = \sum^K_{c=1} \frac{1}{c!} \sum^{c-1}_{j=0}(-1)^{j} {c \choose j}(c-j)^K, \label{eqn:stirling}
\end{eqnarray}
which is the Stirling number of the second kind as shown by \citet{marx:1962} and \citet{salmeri:1962}.
In the 2018 FRC Houston and Detroit championships, $S_K $ is about $1.67\times 10^{69}$ for $K=67$ and $3.66\times 10^{70}$ for $K=68$.
{In this study, two effective estimation procedures are developed to avoid the computational complexity and cost associated with the selection of an appropriate model from a huge number of possible model candidates indexed by the combinations of variant numbers of clusters and clusters of robots.}

{We first propose} the following estimation procedure (Method 1) for model (\ref{eqn:linearcl}):
\begin{description}
\item Step 1. Fit a regression model in (\ref{eqn:linear}) (e.g. the OPR and WMPR models) and compute the LSE $\widehat{\beta}$ of $\beta$.
\item Step 2. Perform {the centroid linkage clustering method of \cite{sokal:1958}} on $\widehat{\beta}$ and obtain cluster estimators $\widehat{g}=  ( \widehat{g}_1, \dots , \widehat{g}_K)^{\top}$ with $\widehat{g}_i$'s  $\in \{ 1, \dots , c\}$.
\item Step 3. Fit a regression model $Y = X^{c} \beta^{c}+ \varepsilon$ and compute the LSE $\widehat{\beta}^{c}$ of $\beta^{c}$, $c = 1 , \dots , K$.
\item Step 4. Compute an estimate $\widehat{\text{AC}}(c)$ of $\text{AC}(c)$ (or an estimate $\widehat{\text{MSPE}}(c)$
of $\text{MSPE}(c)$) and derive the maximizer $\widehat{c} = \arg\max_c \widehat{\text{AC}}(c)$ (or the minimizer $\widehat{c}^{*} =\arg\min_c \widehat{\text{MSPE}}(c)$).
\item Step 5. Estimate $(c_0, g , \beta^{c_0})$ by $(\widehat{c}, \widehat{g} , \widehat{\beta}^{\widehat{c}})$.
\end{description}

The reason of using $\widehat{\beta}$ as an initial estimator is mainly based on the validity of model (\ref{eqn:linear}), which is an overparameterized version of the proposed model for $K>c_{0}$, and the consistency of $\widehat{\beta}$ to $\beta$. As shown by \citet{portnoy:1984}, its convergence rate is the square root of the ratio of the number of robots to the number of observations.
Owing to the constant strength of robots in the same cluster, i.e. $\beta_{i}$'s $\in\{\beta^{c_{0}}_{1},\dots,
\beta^{c_{0}}_{c_{0}}\}$, the centroid linkage clustering method{, which is one of the hierarchical clustering methods {\citep{gordon:1987},}} is reasonably used to measure the distance between any two clusters. 
For $c\geq c_{0}$, it is further ensured that $\widehat{\beta}^{c}$, which is a consistent estimator of $\beta^{c}$ with $\beta^{c}_{i}$'s $\in\{\beta^{c_{0}}_{1},\dots,
\beta^{c_{0}}_{c_{0}}\}$, will be more and more precise as $c$ decreases to $c_{0}$.  
Thus, the determination of $c_{0}$ can be naturally transferred to the model selection problem in such a dimension reduction of the parameter space.
In the above estimation procedure, the LSE of $\beta^{1}_{0}$ is directly derived as $\widehat{\beta}^{1}=\sum^{M}_{s=1}(Y^{\text{R}}_{s}+Y^{\text{B}}_{s})/6M$ for the OPRC model and $\widehat{\beta}^{1}=0$ for the WMPRC model.
Different from existing criteria \citep[e.g.][]{krzanowski:1985, tibshirani:2001, sugar:2003, wang:2010} in the cluster analysis, the optimal number of clusters $c_{0}$ is determined by either maximizing the leave-one-match-out cross-validation estimate of the accuracy $\text{AC}(c)$ or minimizing the leave-one-match-out cross-validation estimate of the mean square prediction error $\text{MSPE}(c)$ with respect to the number of clusters $c$, where the left-out match (testing data) plays as the role of a future run and the remaining matches (training data) are used to build up competing models.
Although the properties $\text{AC}(c_0)>\text{AC}(c)+O(M^{-1})$ and $\text{MSPE}(c_0)<\text{MSPE}(c)+O(M^{-1})$
hold for $c<c_0 \ll K$, the cross-validation criterion, which is akin to the Akaike information criterion (AIC) {\citep{akaike:1974},} is inconsistent in model selection because of the properties $\text{AC}(c)=\text{AC}(c_{0})+O(M^{-1})$ and $\text{MSPE}(c)=\text{MSPE}(c)+O(M^{-1})$
for $c_0<c \ll K$. However, in our application,
the imprecise estimates of robot strengths in a set of nested models of the form (\ref{eqn:linearcl1}) are found to result in 
$\widehat{\text{AC}}(c_{1})>\widehat{\text{AC}}(c_{2})$ for $c_{2}>c_{1}\geq\widehat{c}$ and $\widehat{\text{MSPE}}(c_{1})<\widehat{\text{MSPE}}(c_{2})$ for $c_{2}>c_{1}\geq\widehat{c}^{*}$.
This implicitly implies that $\widehat{c}$ and $\widehat{c}^{*}$ should be much smaller than $K$ and close to $c_{0}$. A more thorough study warrants future research.
It is notable that 
$\widehat{\text{AC}}(c)$ and $\widehat{\text{MSPE}}(c)$ are not only used to assess the predictive ability of models, but also to estimate the number of clusters
$c_0$ in model (\ref{eqn:linearcl}).

In fact, clusters of robots $\widehat{g}_i$'s can be obtained by performing the cluster analysis on the estimator $\widetilde{\beta}^{(c+1)}$ sequentially with $c =K-1 , \dots , 2$.
As an alternative of Method 1, the second estimation procedure (Method 2) is further proposed as follows:
\begin{description}
\item Step 1. Fit a regression model $Y = X \beta+ \varepsilon$ and compute the LSE $\widehat{\beta}$ of $\beta$ and 
$\widehat{\text{AC}}(K)$ (or $\widehat{\text{MSPE}}(K)$).
\item Step 2. Perform {the centroid linkage clustering method} on $\widetilde{\beta}^{K}\stackrel{\triangle}{=}\widehat{\beta}$ and obtain cluster estimators $\widehat{g}=  \big(\widehat{g}_1, \dots , \widehat{g}_K\big)^{\top}$ with $\widehat{g}_{i}$'s  $\in \{ 1, \dots , K-1\}$.
\item Step 3. Fit a regression model $Y = X^{K-1} \beta^{K-1} + \varepsilon$ and compute the LSE $\widehat{\beta}^{K-1}$ of $\beta^{K-1}$ and $\widehat{\text{AC}}(K-1)$ (or $\widehat{\text{MSPE}}$ $(K-1)$).
\item Step 4. Repeat Steps 2--3 for $\widehat{g}$, $\widetilde{\beta}^{c}$, $\widehat{\beta}^{c-1}$, and $\widehat{\text{AC}}(c-1)$ (or $\widehat{\text{MSPE}}(c-1)$), $c = K-1, \dots , 2$.
\item Step 5. Estimate $c_0$, $g $, and $\beta^{c_0}$ by $\widehat{c} = \arg\max_c \widehat{\text{AC}}(c)$ (or $\widehat{c}^{*} = \arg\min_c \widehat{\text{MSPE}}(c)$), $\widehat{g}$, and $\widehat{\beta}^{\widehat{c}}$, respectively.
\end{description}
As we can see, clusters of robots $\widehat{g}_i$'s are obtained by performing cluster analysis on $\widetilde{\beta}^{K}$ in Method 1 but on $\widetilde{\beta}^{c+1}$, $c = K-1, \dots , 2$, in Method 2. It is notable that Method 2 shares the asymptotic properties of Method 1 because this estimation procedure uses the same consistent initial estimator. Although both estimation procedures might produce different estimated numbers of clusters, their estimated accuracies for the proposed model are close to each other in our application.

In the proposed model (\ref{eqn:linearcl}), the
multiple comparison procedures such as those proposed by \citet{hochberg:1987}, \citet{hsu:1996}, and \citet{hothorn:2008},
are possible avenues for the determination of the number of clusters and  clusters of robots. However, how to control the overall type I error rate for the simultaneous inference of $\{\beta_{i}-\beta_{j}:i \neq j\}$ is still a challenging task. Furthermore, such an inference procedure might not be able to achieve the prediction purpose. As indicated in the introduction, a small ratio of the number of observations to the number of robots will affect the precision of the LSE $\widehat{\beta}$
of $\beta$ in (\ref{eqn:linear}). It tends to choose a small number of clusters and, hence, lead to poor prediction on match outcomes. 

\begin{rem}
The formulation in (\ref{eqn:linearcl1})  for the WMPRC model can be rewritten as
$Y = \bar{X}^{c} \beta^{*c} + \varepsilon$, where $\bar{X}^{c}$ is defined as $\bar{X}$ with the $j$th column being $X^{c}_{(j)}-(n_{1}/n_{c})X^{c}_{(c)}$, 
and 
$\beta^{*c}=\big(\beta^{c}_{1},\dots, \beta^{c}_{c-1}\big)^{\top}$ with $\beta^{c}_{c}=-\sum^{c-1}_{j=1}n_{j}\beta^{c}_{j}/n_{c}$ and $n_{j}=\sum^{n}_{i=1}I(g_i = j), j=1,\dots, c$.
As in the context of regression analysis, the LSE of $\beta^{c}$ in the OPRC model is
\begin{eqnarray}
 \widehat{\beta}^{c}=\big(X^{c\top}X^{c}\big)^{-1}X^{c\top}Y\label{eqn:LSE_OPRC}
 \end{eqnarray}
and the LSE of $\beta^{c}$ in the WMPRC model is
\begin{eqnarray}
\widehat{\beta}^{c}=\begin{pmatrix} \widehat{\beta}^{*c}\\ \widehat{\beta}^{c}_{c}\end{pmatrix}\text{ with }
\widehat{\beta}^{*c}=\big(\bar{X}^{c\top}\bar{X}^{c}\big)^{-1}\bar{X}^{c\top}Y \text{ and }
\widehat{\beta}^{c}_{c}=\frac{-\sum^{c-1}_{j=1}n_{j}\widehat{\beta}^{*c}_{j}}{n_c}\text{ for } c\geq 2.\label{eqn:LSE_WMPRC}
\end{eqnarray}
\qed
\end{rem}

\begin{rem}
Although the Bayesian information criterion (BIC) {\citep{schwarz:1978}} has been widely used in model selection, \citet{giraud:2015} showed that the BIC is valid only when $M$ is much larger than $K$, which is not the case for this scenario. Furthermore, the BIC is infeasible in our setup with the {lack of a particular distribution} assumption on $Y_{s}$ and $D_{s}$, $s=1,\dots, M$. As we can see, existing methods in the latent cluster analysis 
\citep[]{lazarsfeld:1968,goodman:1974,collins:2010} are inapplicable to the proposed model because of their focus mainly on identifying the underlying clusters of $Y_{s}$'s rather than those of $\beta_{i}$'s.
Furthermore, when $c$ is greater than 1, there exists a huge number of possible class membership combinations 
$\big( \sum^{c-1}_{j=0}(-1)^{j} {c \choose j}(c-j)^K\big)/c!$ for $g_{i}$'s  in the perspective of latent class analysis. For different combinations of $g_{i}$'s, it also needs to carry out a complicated computational task in deriving the corresponding design covariates $X^{c}_{(j)}$'s and estimates $\widehat{\beta}^{c}_{j}$'s.
\qed
\end{rem}

\subsection{Agreement {Assessment}}
{Let $R_{0}$ be the collection of FRC ratings of all, playoff, or FRC top-8 robots; $K_{0}$ the size of $R_{0}$; and $\widetilde{\beta}_{0}$ the estimated robot strengths of robots in $R_0$.
To assess the agreement between FRC ratings and estimated robot strengths, the rank correlation, which was first proposed by \citet{han:1987}, of
$R_{0}$ and $\tilde{\beta}_{0}$ is computed as
\begin{eqnarray}
\text{RC}(R_{0},\tilde{\beta}_{0})&=&\frac{1}{K_{0}(K_{0} -1)} \sum \sum_{i \neq j} I \big( \text{sgn}(R_{i} - R_{j}) \cdot \text{sgn}(\widetilde{\beta}_{i} - \widetilde{\beta}_{j}) > 0\big) \nonumber\\
&&+ 0.5 I \big( \text{sgn}(R_{i} - R_{j}) \cdot \text{sgn}(\widetilde{\beta}_{i} - \widetilde{\beta}_{j}) = 0\big), \label{eqn:rankcor}
\end{eqnarray}
where $R_{i}$'s and $\widetilde{\beta}_{i}$'s are the corresponding elements of $R_{0}$ and $\widetilde{\beta}_{0}$.} In fact, this rank-based measure is particularly useful to investigate the monotonic association of two ordinal scale measurements. Other measures such as Kendall's $\tau$ \citep{kendall:1938} and spearman's $\rho$ \citep{spearman:1904} can also be used to assess the agreement between $R_{0}$ and $\widetilde{\beta}_{0}$.

As in the contexts of pattern recognition and information retrieval,  we assess the agreement between the FRC top-8 robots and model-based top-N robots, which are rated by estimated robot strengths, by the precision measure
and the recall measure from \citet{perry:1955}.
Let $\text{RS}_{\text{top}}$ and $\text{RS}(N)$ stand for the corresponding sets of FRC top-8 robots and
model-based top-$N$ robots. The precision and recall measures of $RS_{top}$ and $RS(N)$ are further given by
{
\begin{equation}
\text{Pr}(\text{RS}_{\text{top}},\text{RS}(N)) = \frac{|\text{RS}_{\text{top}} \cap \text{RS}(N)|}{N}\text{ and }
\text{Re}(\text{RS}_{\text{top}},\text{RS}(N)) = \frac{|\text{RS}_{\text{top}} \cap \text{RS}(N)|}{8}.\label{eqn:PRRE}
\end{equation}}
In diagnostic tests, the precision and recall are termed as the positive predictive value per \citet{vecchio:1966} and sensitivity per \citet{yerushalmy:1947}, respectively.
It is notable that the sensitivity is a measure of the intrinsic accuracy of a test whereas the positive predictive value is only a function of the accuracy and prevalence.

\subsection{Stability {Assessment}}

{In this study, we also assess how many matches are needed in the qualification stage to produce a clear picture about robot strengths. This is mainly to make a recommendation on an optimal number of matches in future tournaments to improve planning and logistics. Let $Y^{(t)}_{[\ell]}$ and $X^{(t)}_{[\ell]}$ be the corresponding vector and matrix formed by the first $M_{\ell}$ elements of $Y^{(t)}$ and the first $M_{\ell}$ rows of $X^{(t)}$
with $(t)$ being either $\text{B}$ or $\text{R}$ and $M_{\ell}=\lceil \frac{\ell \times K}{6} \rceil $, $\ell = 6, \dots , m_0$.
Based on data $\big\{ Y^{\text{B}}_{[\ell]}, Y^{\text{R}}_{[\ell]}, X^{\text{B}}_{[\ell]} ,X^{\text{R}}_{[\ell]} \big\}$ and the estimators
$(\widehat{c},\widehat{g})$ (or $(\widehat{c}^{*},\widehat{g})$) of $(c_0, g)$, we define
$\widetilde{\beta}_{[\ell]} $ as the robot strengths obtained from the LSE of $\beta^{c_{0}}$ in model (\ref{eqn:linearcl}) and $\bar{\beta}_{[\ell]}$ as the top-8 robot strength estimates according to the values in $\widetilde{\beta}_{[\ell]}$, $\ell = 6, \dots , m_0$.}

Same with the formulations of $\widehat{\text{AC}}(c)$ 
and $\widehat{\text{MSPE}}(c)$ in (\ref{eqn:PRcv}),
$\widehat{\text{AC}}_{[\ell]}(\widehat{c})$ and $\widehat{\text{MSPE}}_{[\ell]}(\widehat{c}^{*})$
are computed based on data $\big\{ Y^{\text{B}}_{[\ell]}, Y^{\text{R}}_{[\ell]}, X^{\text{B}}_{[\ell]} ,X^{\text{R}}_{[\ell]} \big\}$ and $\widetilde{\beta}_{[\ell]}$, $\ell = 6, \dots , m_0$.
The stability of estimated robot strengths and accuracies can be assessed through $\{ \text{RC}(\widetilde{\beta}_{[\ell]},\widetilde{\beta}_{[\ell+1]}): \ell = 6,\dots , (m_0-1)\}$ (or $\{ \text{RC}(\bar{\beta}_{[\ell]},\bar{\beta}_{[\ell+1]}): \ell = 6,\dots , (m_0-1)\}$) and $\{ \widehat{\text{AC}}_{[\ell]}(\widehat{c}): \ell = 6,\dots , m_0\}$ (or $\{ \widehat{\text{MSPE}}_{[\ell]}(\widehat{c}^{*}): \ell = 6,\dots , m_0\}$), respectively, where the rank correlation function $\text{RC}(\cdot,\cdot)$ is defined in (\ref{eqn:rankcor}). Given these measures and
a pre-determined tolerance threshold, we can further determine an appropriate number of matches in the qualification stage.

\section{An Application to the 2018 FRC Championships} \label{sec:num-example}

\begin{figure}[htbp]
\centering
 \includegraphics[width=7cm,height=9cm]{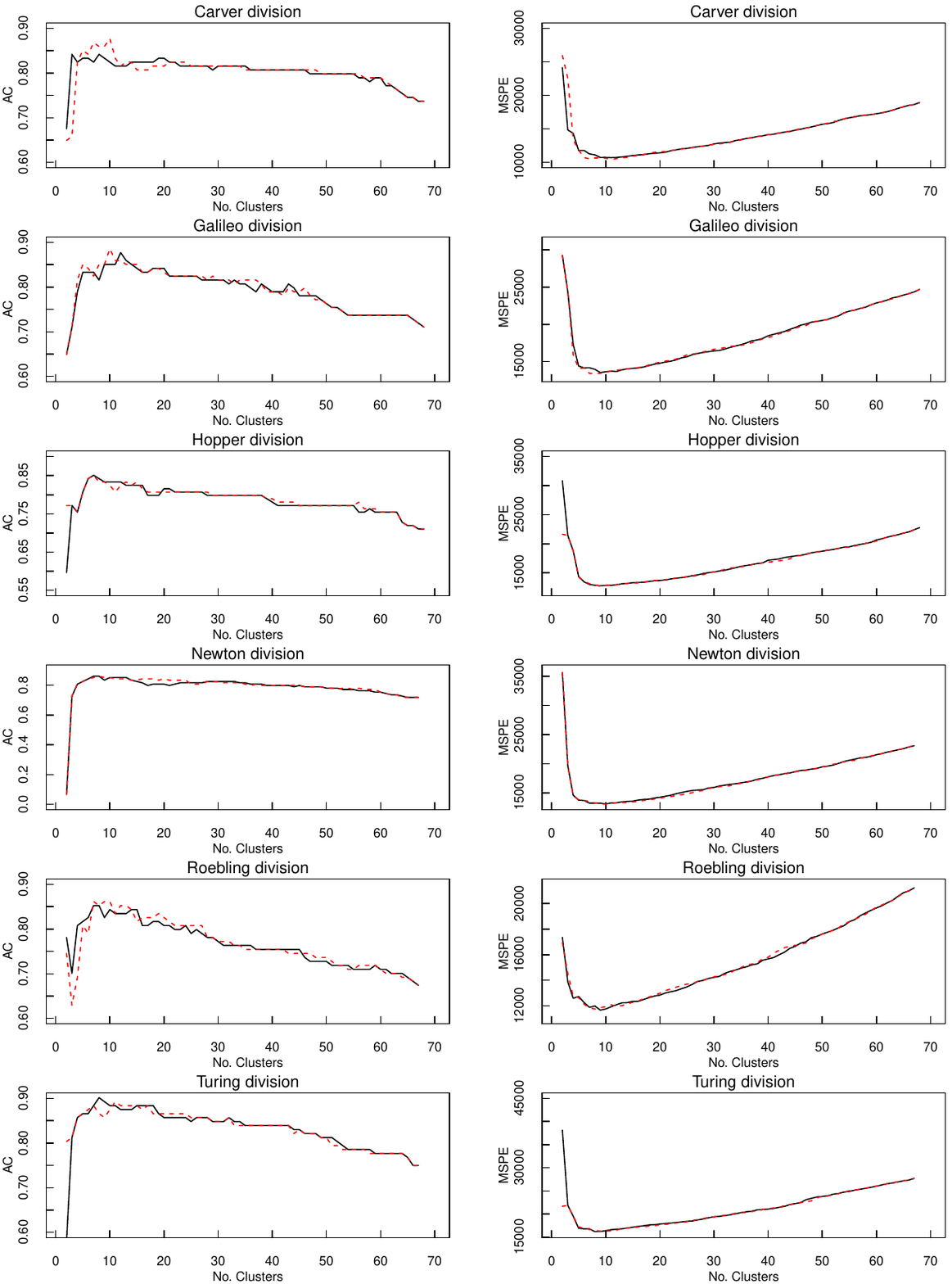}
  \includegraphics[width=7cm,height=9cm]{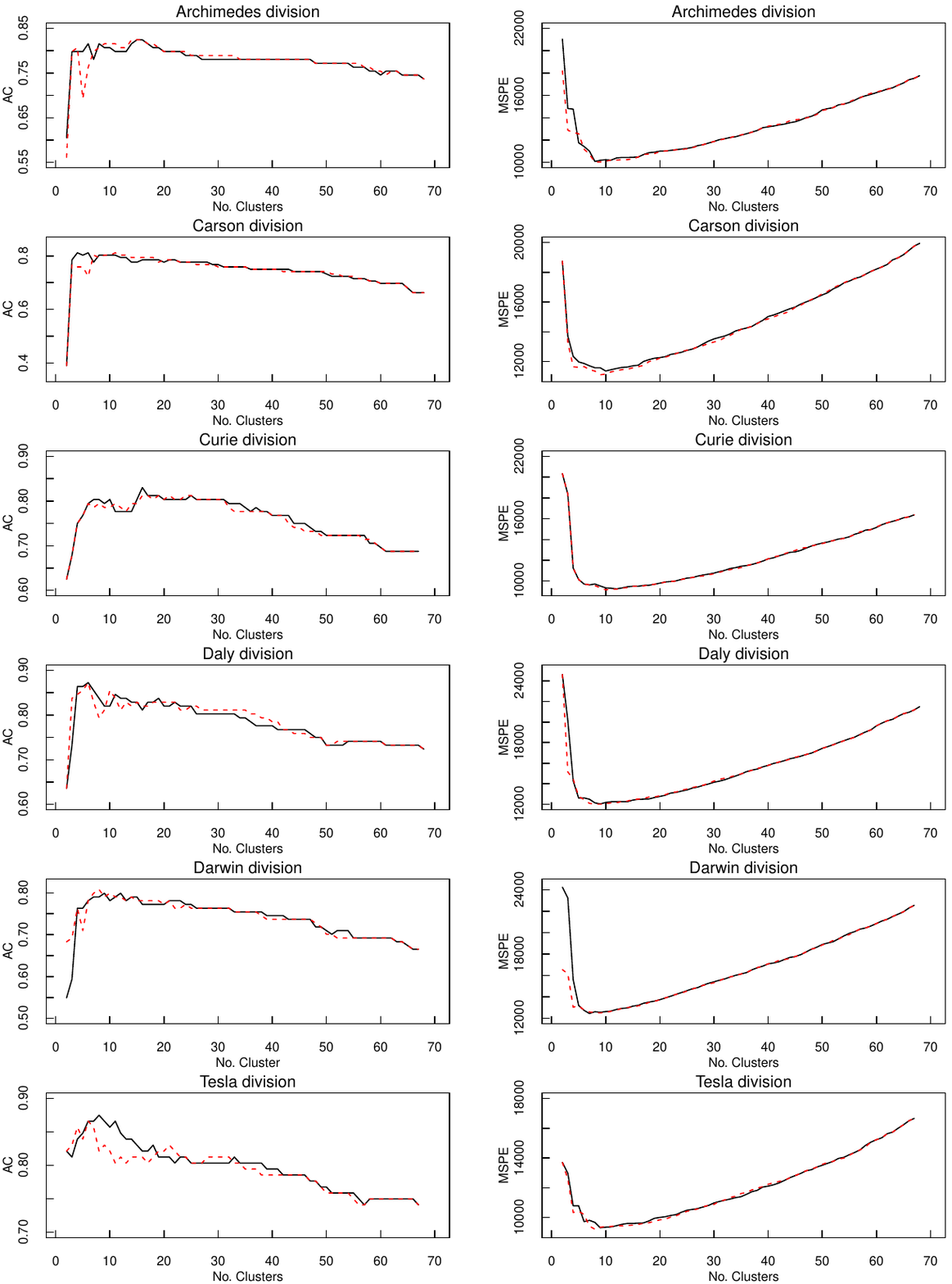}
 \caption{{\footnotesize The estimated accuracy and mean square prediction error curves from the OPRC1 (black-solid line) and OPRC2 (red-dashed line) estimation procedures for the divisions of the 2018 FRC Houston and Detroit championships.}} \label{OPRC_HoustonDetroit} 
  \end{figure}

\begin{figure}[htbp]
\centering
 \includegraphics[width=7cm,height=9cm]{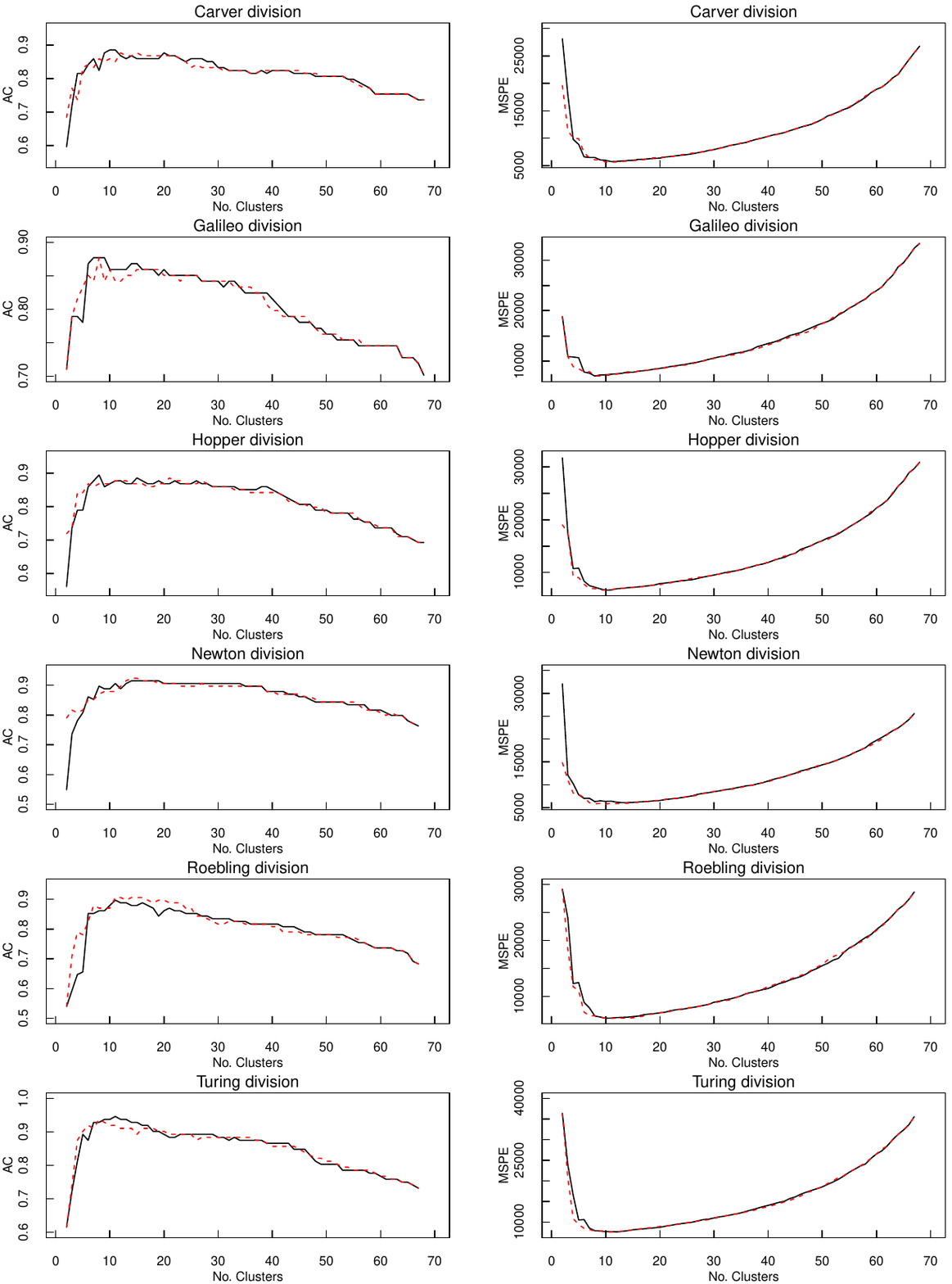}
  \includegraphics[width=7cm,height=9cm]{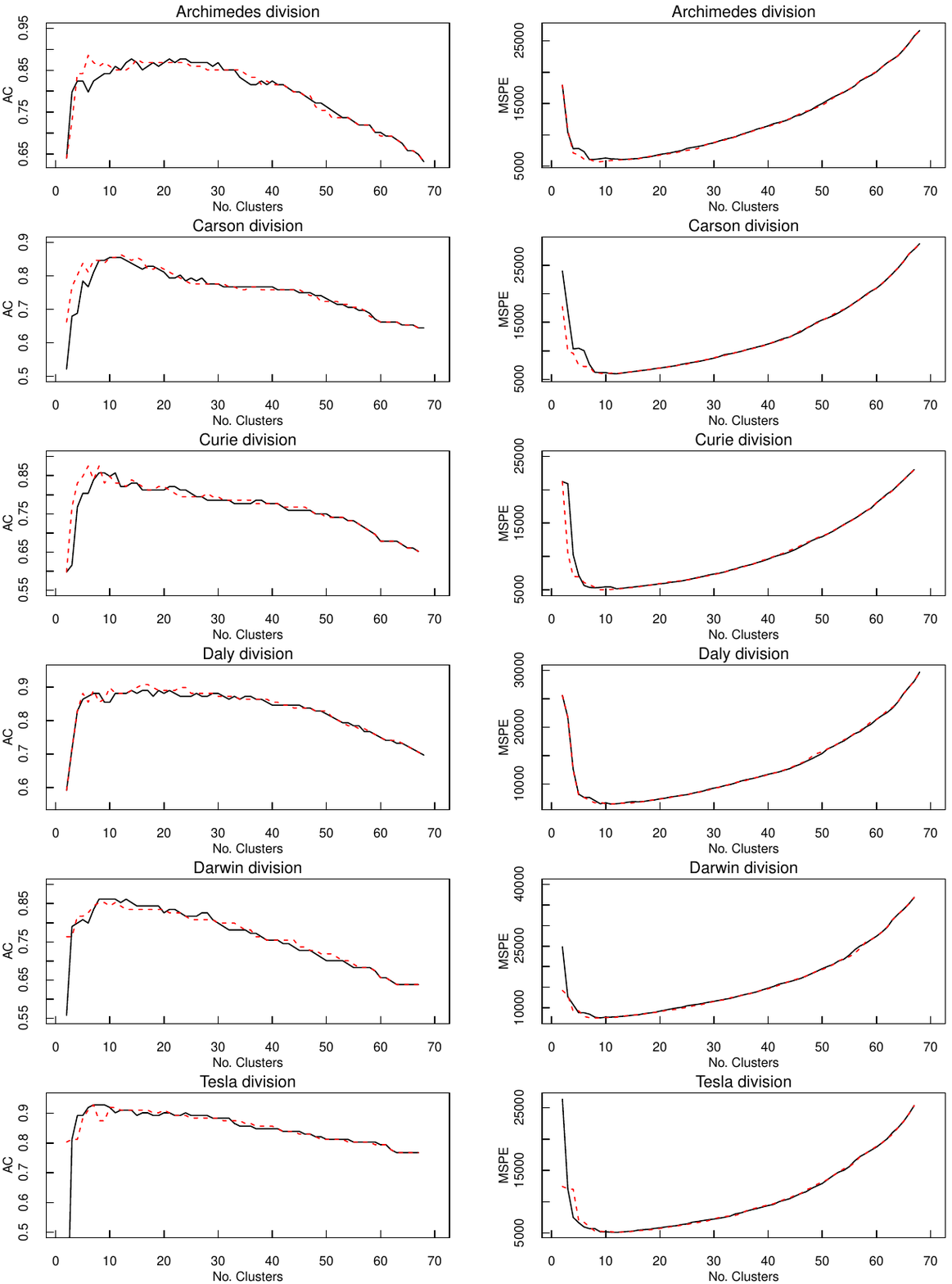}
  \caption{{\footnotesize The estimated accuracy and mean square prediction error curves from the WMPRC1 (black-solid line) and WMPRC2 (red-dashed line) estimation procedures for the divisions of the 2018 FRC Houston and Detroit championships.}} \label{WMPRC_HoustonDetroit} 
\end{figure}

In this section, the OPR and WMPR models in Sections 2.2 and 2.4--2.5 and the OPRC and WMPRC models in Section 3.1 are applied to the 2018 FRC Houston and Detroit championships. As discussed in Remark 3, the WMPRR model is the WMPR model with random robot strengths. Its flaws are also investigated in this application. As of the 2018 competitions, each championship site is composed of six divisions: Carver, Galileo, Hopper, Newton, Roebling, and Turing in Houston; and Archimedes, Carson, Curie, Daly, Darwin, and Tesla in Detroit.
For the determination of the number of clusters and clusters of robots, the developed estimation procedures (Method 1, Method 2) in Section 3.3 are denoted by (OPRC1,OPRC2) for the OPRC model and (WMPRC1,WMPRC2) for the WMPRC model. To avoid verbosity, the corresponding estimation procedures for the OPR and WMPR models are hereinafter denoted by OPR and WMPR estimation procedures.
In Table \ref{tbl:table1}, we summarize
the number of matches and the number of robots of each division in the qualification stage and the playoff stage.
Table \ref{tbl:table2} further displays the estimated numbers of clusters from the OPRC1, OPRC2, WMPRC1, and WMPRC2 estimation procedures. 
Apparently, the estimated numbers of clusters are relatively small compared to the number of robots. 
Moreover, it is also shown in Fig. \ref{OPRC_HoustonDetroit} and Fig. \ref{WMPRC_HoustonDetroit} that there exist an inverted U-shaped relationship between the estimated accuracy and the number of clusters and a U-shaped relationship between the estimated mean square prediction error and the number of clusters.

\begin{table}[htbp]
\centering
 \caption{{\footnotesize The number of matches $M$ and the number of robots $K$ in the qualification stage and the number of matches $M^{*}$ and the number of robots $K^{*}$ in the playoff stage.}} \label{tbl:table1}
\includegraphics[width=6.5cm, height=3cm]{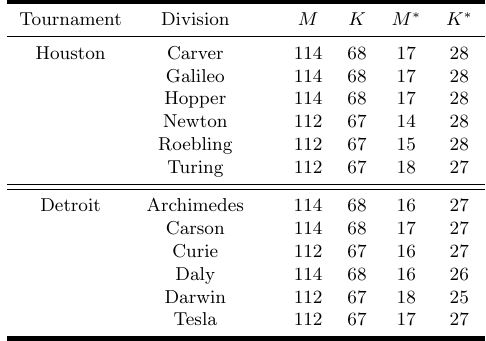}
\end{table}

\begin{table}[htbp]
  \centering
  \caption{{\footnotesize The estimated numbers of clusters $\widehat{c}$ and $\widehat{c}^{*}$, in which the numbers within the brace are the maximum and minimum cluster sizes, of $c_{0}$ from the OPRC1, OPRC2, WMPRC1, and WMPRC2 estimation procedures on qualification data.}} \label{tbl:table2}
\includegraphics[width=12cm,height=3.5cm]{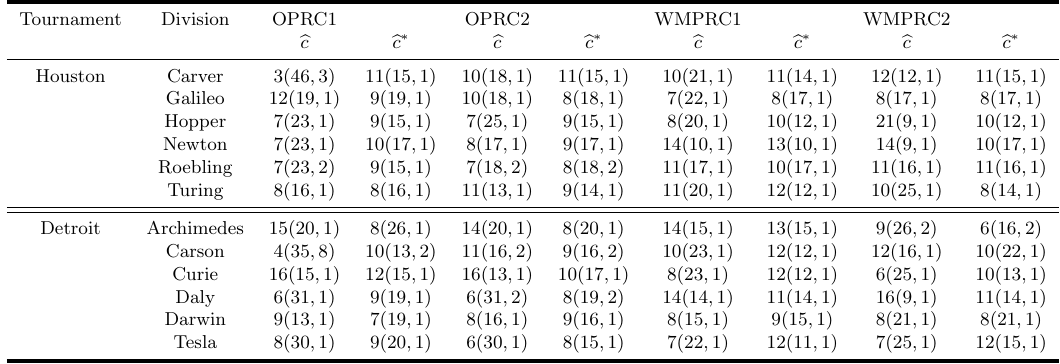}
\end{table}

Since the  research interest mainly focuses on predicting match outcomes, our investigation is based on the estimated number of clusters $\widehat{c}$.
In the considered twelve divisions,
the corresponding ratios of the number of observations to the estimated number of clusters range from 22.33 to 45.33 for the WMPRC model and
from 6.48 to 22.67 for the WMPRC model.
With qualification data, Table \ref{tbl:table3} displays the estimated accuracies of the form in (\ref{eqn:PRcv}) from different estimation procedures for the divisions of {the} Houston and Detroit tournaments. {For the accuracies of the investigated models, approximate $0.95$-confidence intervals, which are constructed as in (\ref{eqn:ci}), are further provided in Table \ref{tbl:table3}. Although the accuracies from each estimation procedure are not significantly different for all divisions of a tournament, there is no strong evidence to aggregate the results across divisions to obtain a more robust estimator of the accuracy. This is mainly because the underlying models in some divisions might be completely different.}
The overall estimated accuracies from the {AS,} OPR, OPRC1, OPRC2, WMPR, WMPRC1, WMPRC2, and WMPRR estimation procedures are computed as {$70.3\%$,} $71.0\%$, $85.0\%$, $85.2\%$, $69.5\%$, $89.0\%$, $89.3\%$, and $70.1\%$. {Based on the test statistic in (\ref{eqn:test}), it is shown that the accuracies of the OPRC and WMPRC models are significantly higher than those of the OPR and WMPR models in all divisions (p-values $\leq 0.007$). As we can see, the accuracies of the AS, OPR, and WMPR models are not significantly different in most divisions (p-values $\geq 0.127$), except that the accuracy of the WMPR model is significantly lower than the accuracies of the AS and OPR models in {the} Archimedes division (p-values $=0.003$ and $0.007$), the accuracy of the OPR model in {the} Daly division (p-value $=0.014$), and the accuracy of the AS model in {the} Darwin division (p-value $=0.029$).} The performance of the WMPRR model is only comparable with that of the WMPR model. Its poor prediction is probably caused by over-simplifying the heterogeneous feature of robot strengths.
The performance of the OPRC1 and OPRC2 estimation procedures {and that of the WMPRC1 and WMPRC2 estimation procedures
are further found to be comparable in all divisions (p-values $\geq 0.158$)}. For the divisions of {the} Houston and Detroit tournaments, the estimated accuracies of the OPRC and WMPRC models are about $8\%-19\%$ and $14\%-26\%$ higher than those of the OPR and WMPR models, respectively. 
{Although the predictive ability of the WMPRC model is generally better than that of the OPRC model, their accuracies are not significantly different in all divisions (p-values $\geq 0.079$).} Despite using a smaller data set, it is possible for the WMPRC model to produce a more meaningful prediction compared to the OPRC model, which does not take into account the feature of paired comparisons. Although the estimated numbers of clusters from the WMPRC1 and WMPRC2 estimation procedures are different in most of the divisions, the predictive capacities of both models are close to each other. The same conclusion can be drawn for the OPRC1 and OPRC2 estimation procedures.
As emphasized in Section 3.3, the accuracy and mean square prediction error tend to overestimate the corresponding numbers of clusters in the OPRC and WMPRC models.

\begin{table}[htbp]
  \centering
  \caption{{\footnotesize  The estimated accuracies {(approximate $0.95$-confidence intervals)} from the {AS,} OPR, OPRC1, OPRC2, WMPR, WMPRC1, WMPRC2, and WMPRR estimation procedures on qualification data.}} \label{tbl:table3}
\includegraphics[width=12cm,height=3.5cm]{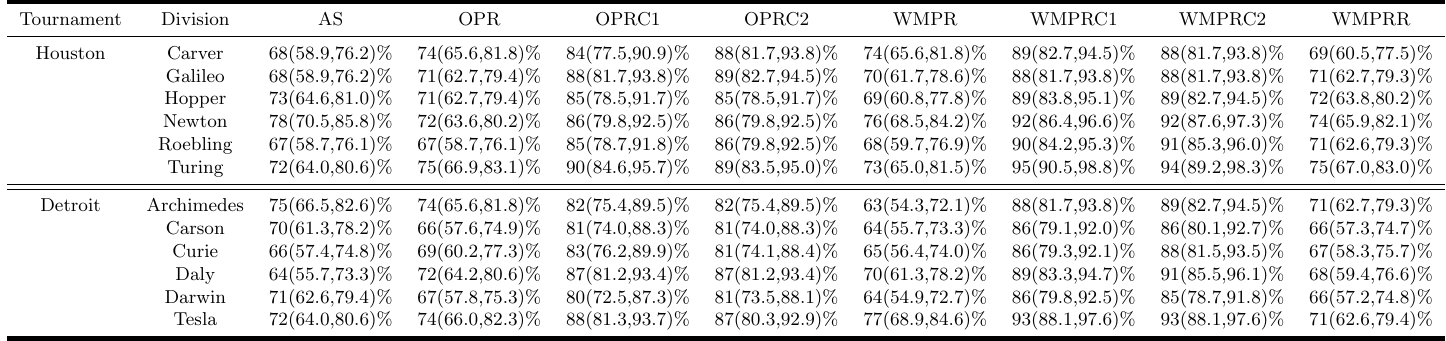}
\end{table}

{Provided that model (\ref{eqn:linearcl}) is valid for all divisions in the same tournament, we can add up their estimated accuracy functions in (\ref{eqn:PRcv}) to estimate the number of clusters $c_{0}$. The estimated numbers of clusters from the OPRC1, OPRC2, WMPRC1, and WMPRC2 estimation procedures are 8, 10, 11, and 15, respectively,  in Houston and 6, 10, 11, and 10, respectively, in Detroit.
The estimated accuracies from these estimation procedures are further displayed in Table \ref{tbl:table34}. Compared to the performance of the corresponding estimation procedures with estimated numbers of clusters for each division, our test shows that there is no significant difference in most divisions (p-values $\geq 0.079$). However, our OPRC1, OPRC2, and WMPRC2 estimation procedures significantly outperform the OPRC1, OPRC2, and WMPRC2 estimation procedures, in which a common estimated number of clusters is used for all divisions in the same tournament, in {the} Galileo and Curie divisions (p-values $=0.004$ and $0.051$), {the} Curie and Tesla divisions (p-values $=0.041$ and $0.047$), and {the} Turing and Archimedes divisions (p-value $=0.012$ and $0.041$), respectively. Based on these results, there is no strong evidence to use the same number of clusters for different divisions of a tournament.}
 
 \begin{table}[htbp]
  \centering
  \caption{{\footnotesize {The estimated accuracies (approximate $0.95$-confidence intervals) and accuracies from the OPRC1, OPRC2, WMPRC1, and WMPRC2, estimation procedures, in which a common estimated number of clusters is used for all divisions in the same tournament, on qualification and playoff data, respectively.}}} \label{tbl:table34}
\includegraphics[width=12cm,height=3.5cm]{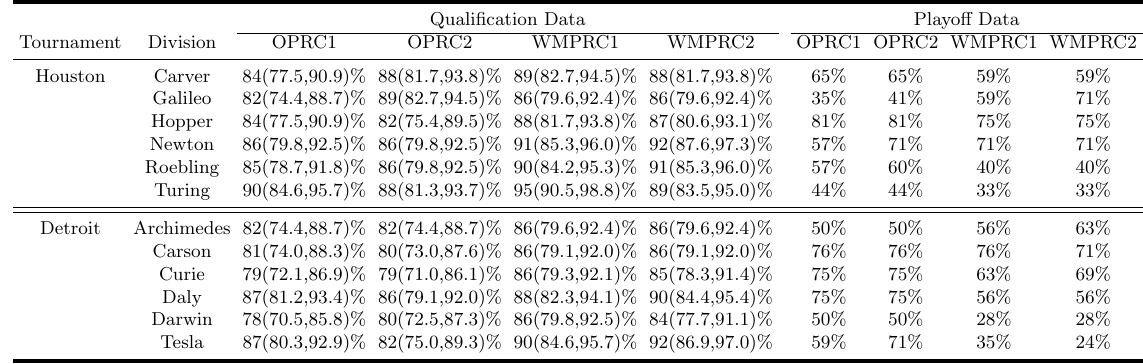}
\end{table}

Since robots are not randomly assigned to matches and alliances in the playoff stage, the effects of some confounders might be ignored in model fitting for qualification data. By treating qualification and playoff data as training and testing data, the accuracy of a test on match outcomes
in the playoff stage is computed as that in (\ref{eqn:PRcvpo}) and is expected to be low (see Table \ref{tbl:table4}). It can seen that the predictive ability of the OPR model is comparable with or even better than that of the OPRC model.
Except in {the} Turing division of {the} Houston tournament and {the} Darwin and Tesla divisions of {the} Detroit tournament, the WMPR model is comparable with or better than the WMPRC model. 
The WMPRR model are further shown to have very poor prediction for match outcomes in the playoff stage. {In addition, the OPR model has similar performance to the AS model except in {the} Roebling division and}
outperforms the WMPR model except in {the} Galileo and Newton divisions of {the} Houston tournament and {the} Archimedes division of {the} Detroit tournament. 
{Applying the OPRC and WMPRC models with a common number of clusters for all divisions in the same tournament, yields the poor predictive capacities found in Table \ref{tbl:table34}.}
For all, playoff, and FRC top-8 robots, the corresponding rank correlations of the form in (\ref{eqn:rankcor}) between FRC ratings and estimated robot strengths from the OPR model are generally higher than those between FRC ratings and estimated robot strengths from the rest models (Tables \ref{tbl:table5}--\ref{tbl:table7}). This particularly explains why the predictive ability of the OPR model is better than those of the OPRC, WMPR, and WMPRC models in the playoff stage.

\begin{table}[htbp]
  \centering
  \caption{{\footnotesize The accuracies from the {AS,} OPR, OPRC1, OPRC2, WMPR, WMPRC1, WMPRC2, and WMPRR estimation procedures on playoff data.}} \label{tbl:table4}
\includegraphics[width=12cm,height=3.5cm]{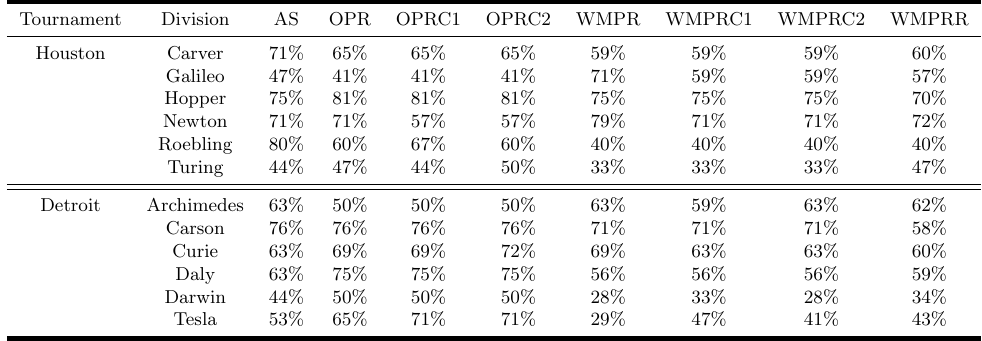}
\end{table}

\begin{table}[htbp]
  \centering
      \caption{{\footnotesize  The rank correlations between FRC ratings and model-based robot strengths, which are estimated by the OPR, OPRC1, OPRC2, WMPR, WMPRC1, and WMPRC2 estimation procedures, of all robots.}} \label{tbl:table5}
\includegraphics[width=12cm,height=3.5cm]{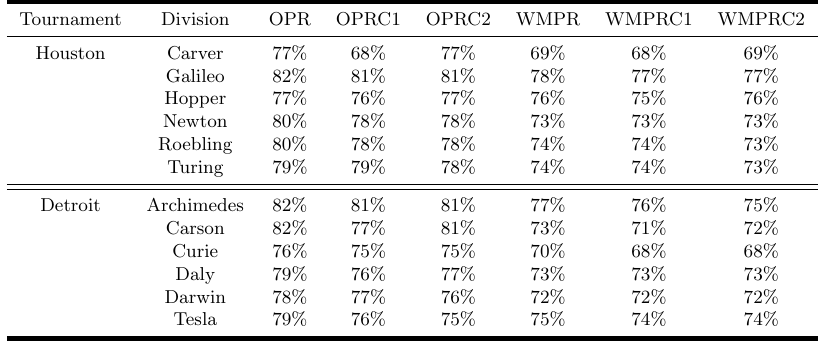}
\end{table}

\begin{table}[htbp]
  \centering
      \caption{{\footnotesize  The rank correlations between FRC ratings and model-based robot strengths, which are estimated by the OPR, OPRC1, OPRC2, WMPR, WMPRC1, and WMPRC2 estimation procedures, of playoff robots.}} \label{tbl:table6}
\includegraphics[width=12cm,height=3.5cm]{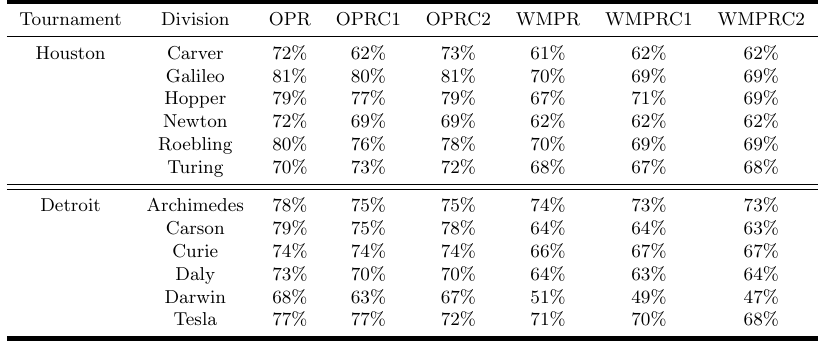}
\end{table}

\begin{table}[htbp]
  \centering
      \caption{{\footnotesize  The rank correlations between FRC ratings and model-based robot strengths, which are estimated by the OPR, OPRC1, OPRC2, WMPR, WMPRC1, and WMPRC2 estimation procedures, of FRC top-8 robots.}} \label{tbl:table7}
\includegraphics[width=12cm,height=3.5cm]{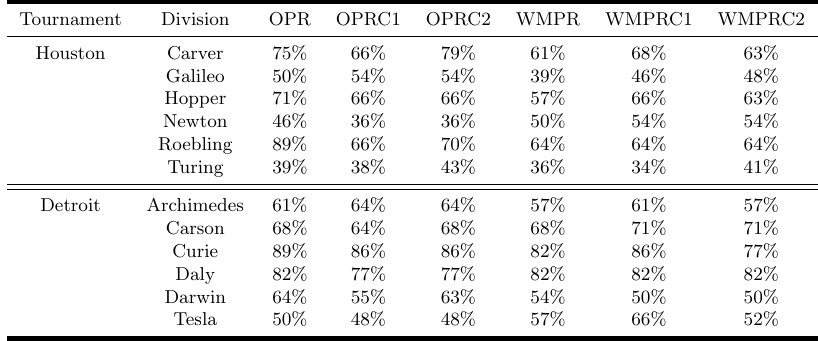}
\end{table}

For the agreement between FRC ratings and model-based robot strengths of all robots in Table \ref{tbl:table5} or playoff robots in
Table \ref{tbl:table6}, 
the rank correlations between FRC ratings and robot strengths estimated by the OPR, OPRC1, and OPRC2 estimation procedures are comparable in all divisions. The same conclusion can be further drawn for
the rank correlations between FRC ratings and robot strengths estimated by the WMPR, WMPRC1, and WMPRC2 estimation procedures.
We further find that the rank correlations between FRC ratings and robot strengths estimated by the OPR, OPRC1, and OPRC2 estimation procedures
are higher than those between FRC ratings and robot strengths estimated by the WMPR, WMPRC1, and WMPRC2 estimation procedures. Generally speaking, there exist very strong monotonic associations among robot strengths estimated by the OPR, OPRC1, and OPRC2 estimation procedures, and among robot strengths estimated by the WMPR, WMPRC1, and WMPRC2 estimation procedures (Table \ref{tbl:table7.1}). Due to a very small estimated number of clusters from the OPRC1 estimation procedure in {the} Carver division of {the} Houston tournament, 
the lower rank correlations are expected between robot strengths estimated by the OPRC1 estimation procedure and robot strengths estimated by the OPR and OPRC2 estimation procedures.
As we can see in Table \ref{tbl:table7.2}, there are 
relatively lower rank correlations 
between robot strengths estimated by the OPR, OPRC1, and OPRC2 estimation procedures, and by the WMPR, WMPRC1, and WMPRC2 estimation procedures.  For the agreement
between FRC ratings and model-based robot strengths of FRC top-8 robots (Table \ref{tbl:table7}), the rank correlation between FRC ratings and robot strengths estimated by the OPR estimation procedure
is comparable with or even higher than the rank correlations between FRC ratings and robot strengths estimated by the OPRC1, OPRC2, WMPR, WMPRC1, and WMPRC2 estimation procedures.
Overall, the rank correlations between FRC ratings and model-based robot strengths are not strong in most of the divisions.

\begin{table}[htbp]
  \centering
      \caption{{\footnotesize  The rank correlations among robot strengths estimated by the OPR, OPRC1, and OPRC2 estimation procedures, and  
among robot strengths estimated by the WMPR, WMPRC1, and WMPRC2 estimation procedures.}} \label{tbl:table7.1}
\includegraphics[width=12cm,height=3.5cm]{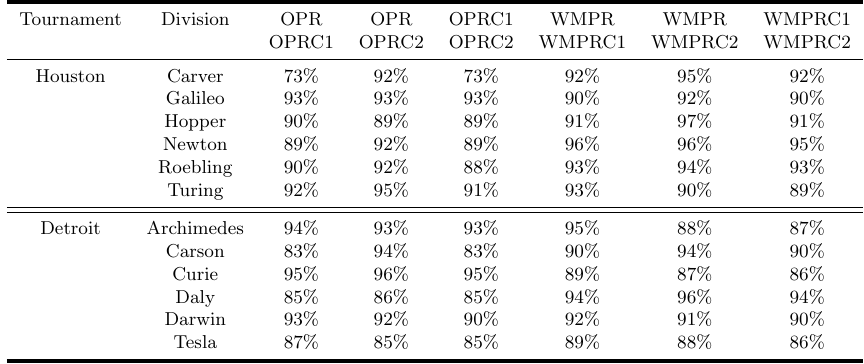}
\end{table}

\begin{table}[htbp]
  \centering
      \caption{{\footnotesize  The rank correlations between robot strengths estimated by the OPR, OPRC1, and OPRC2 estimation procedures and robot strengths estimated by the WMPR, WMPRC1, and WMPRC2 estimation procedures.}} \label{tbl:table7.2}
\includegraphics[width=12cm,height=3.5cm]{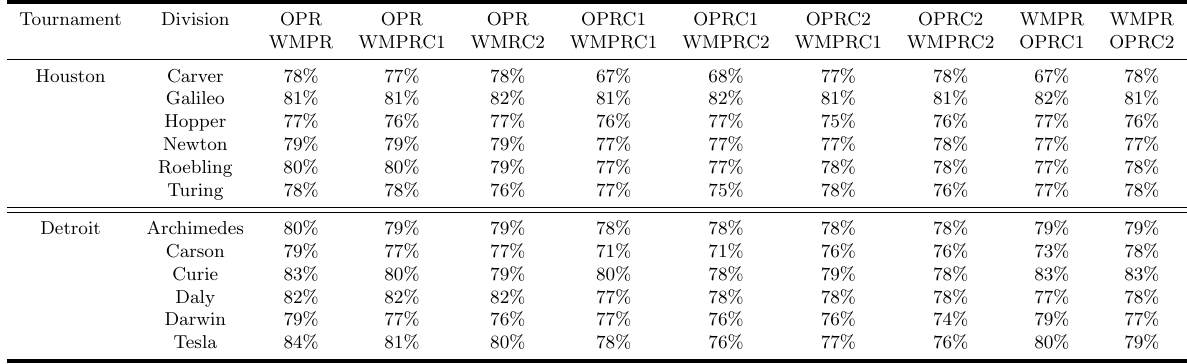}
\end{table}

In Tables \ref{tbl:table8}--\ref{tbl:table9}, it is shown that the precisions of the form in (\ref{eqn:PRRE}) between FRC top-8 robots and OPRC1-based and OPRC2-based top-N robots are comparable with or higher than the precision between FRC top-8 robots and OPRC-based top-N robots, $N=8,16$.
Except for FRC top-8 robots and model-based top-8 robots
in {the} Turing division of {the} Detroit tournament, the precision between FRC top-8 robots and WMPRC1-based top-N robots
is comparable with or higher than the precisions between FRC top-8 robots and WMPR-based and WMPRC2-based top-N robots.
Further, the precisions between FRC top-8 robots and OPRC1-based and OPRC2-based top-N robots are higher than 
the precision between FRC top-8 robots and WMPRC1-based top-N robots
except in {the} Hopper division of {the} Houston tournament. Evidenced by the results in Tables \ref{tbl:table10}--\ref{tbl:table11},
the conclusion for the precision of FRC top-8 robots and model-based top-N robots can also be drawn for
the recall of FRC top-8 robots and model-based top-N robots.

\begin{table}[htbp]
  \centering
      \caption{{\footnotesize Precisions of FRC top-8 robots and model-based top-8 robots, which are determined by the OPR, OPRC1, OPRC2, WMPR, WMPRC1, and WMPRC2 estimation procedures.}} \label{tbl:table8}
\includegraphics[width=12cm,height=3.5cm]{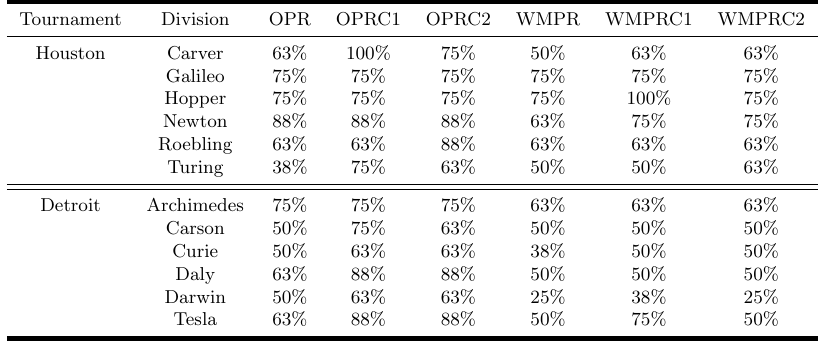}
\end{table}

\begin{table}[htbp]
  \centering
      \caption{{\footnotesize  Precisions of FRC top-8 robots and model-based top-16 robots, which aredetermined by the OPR, OPRC1, OPRC2, WMPR, WMPRC1, and WMPRC2 estimation procedures.}} \label{tbl:table9}
\includegraphics[width=12cm,height=3.5cm]{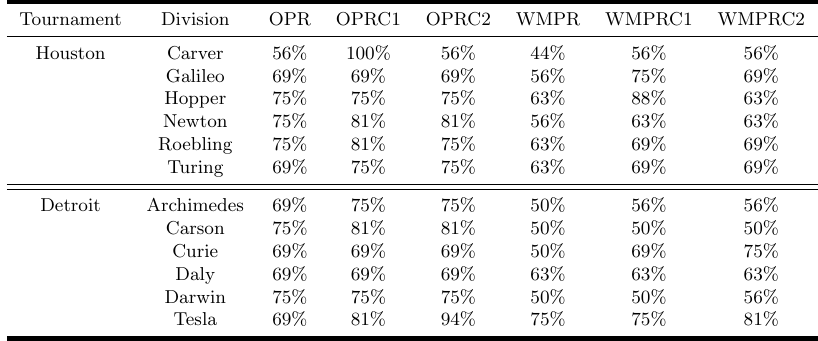}
\end{table}

\begin{table}[htbp]
  \centering
      \caption{{\footnotesize Recalls of FRC top-8 robots and model-based top-8 robots, which are determined by the OPR, OPRC1, OPRC2, WMPR, WMPRC1, and WMPRC2 estimation procedures.}} \label{tbl:table10}
\includegraphics[width=12cm,height=3.5cm]{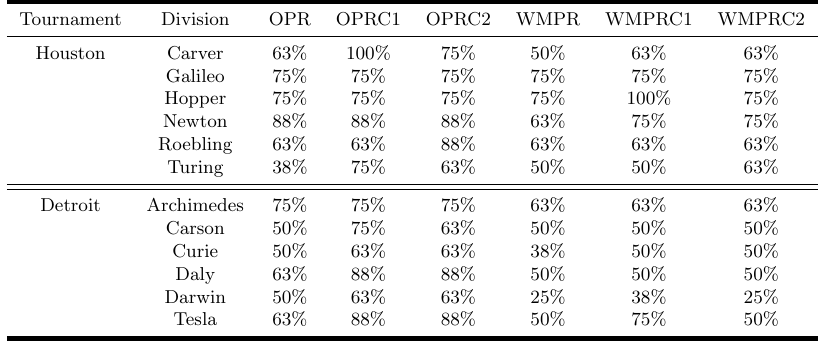}
\end{table}

\begin{table}[htbp]
  \centering
      \caption{{\footnotesize Recalls of FRC top-8 robots and model-based top-16 robots, which are determined by the OPR, OPRC1, OPRC2, WMPR, WMPRC1, and WMPRC2 estimation procedures.}} \label{tbl:table11}
\includegraphics[width=12cm,height=3.5cm]{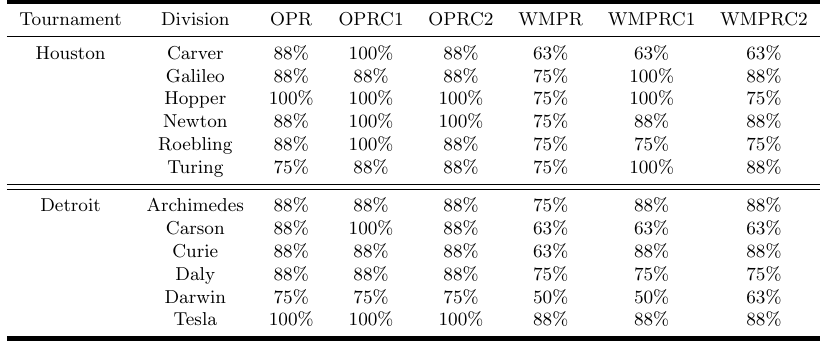}
\end{table}

In Tables \ref{tbl:table12}--\ref{tbl:table13}, we can observe that $M_{9}$ matches should be enough to ensure the stability of accuracies of the OPRC and WMPRC models in the qualification stage. Tables \ref{tbl:table14}--\ref{tbl:table15} further show that the OPRC2, WMPRC1, and WMPRC2 estimation procedures have relatively high rank correlations of estimated robot strengths on $M_{\ell}$ and $M_{\ell+1}$ matches, $\ell=6,\dots, 9$. Except in {the} Carver division of {the} Houston tournament, the same conclusion can be drawn for the OPRC1 estimation procedure. It must be emphasized that these observed results can only refer to the examined case study. 
According to model-based top-8 robots, which are selected according to estimated robot strengths on $M$ matches, high rank correlations of estimated robot strengths on $M_{\ell}$ and $M_{\ell+1}$ matches, $\ell=6,\dots, 9$, (see Tables \ref{tbl:table16}--\ref{tbl:table17}) are found in {the} Archimedes and Tesla divisions from the OPRC1 estimation procedure; in {the} Galileo, Turing, Archimedes, Darwin, and Tesla divisions from the OPRC2 estimation procedure; in {the} Carver, Newton (except for $\ell=6$), Turing, Carson, and Daly divisions from the WMPRC1 estimation
procedure; and in {the} Carver, Hooper, Newton, Carson, and Daly divisions from the WMPRC2 estimation procedure.

\begin{table}[htbp]
  \centering
      \caption{{\footnotesize The accuracies, which are estimated by the OPRC1 and OPRC2 estimation procedures, on $M_{6},\dots,M_{10}$ matches in the qualification stage.}}
      \label{tbl:table12}
\includegraphics[width=12cm,height=3.75cm]{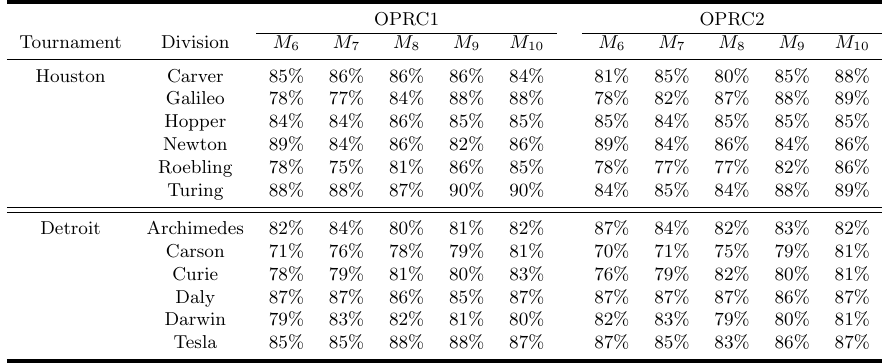}
\end{table}

\begin{table}[htbp]
  \centering
      \caption{{\footnotesize The accuracies, which are estimated by the WMPRC1 and WMPRC2 estimation procedures, on $M_{6},\dots,M_{10}$ matches in the qualification stage.}}
      \label{tbl:table13}
\includegraphics[width=12cm,height=3.5cm]{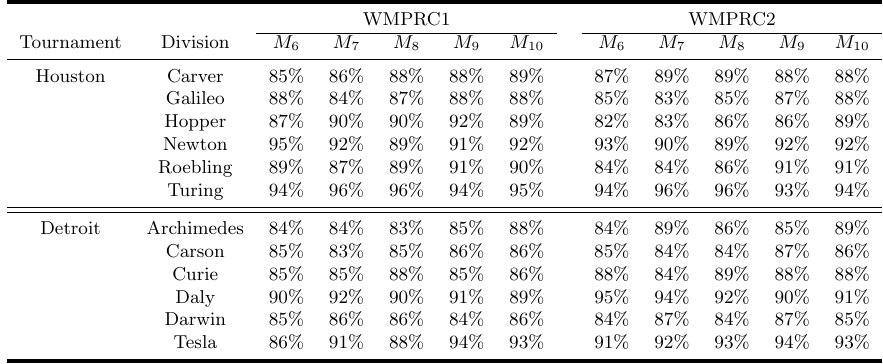}
\end{table}

\begin{table}[htbp]
  \centering
      \caption{{\footnotesize The rank correlations of robot strengths, which are estimated by the OPRC1 and OPRC2 estimation procedures, of all robots on $M_{\ell}$ and $M_{\ell+1}$ matches, $\ell=6,\dots, 9,$ in the qualification stage.}}
      \label{tbl:table14}
\includegraphics[width=12cm,height=3.5cm]{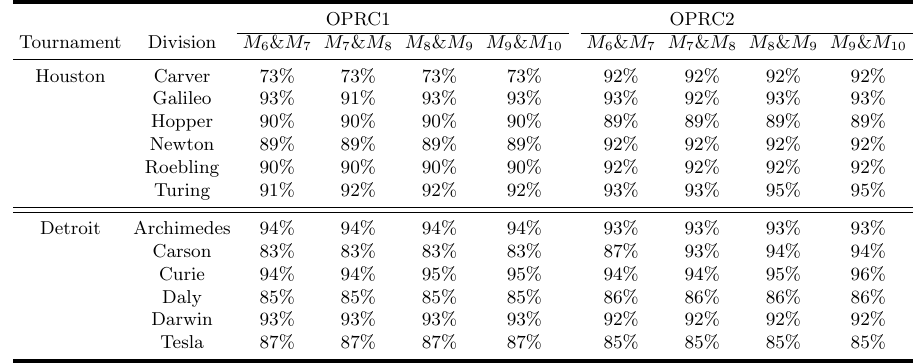}
\end{table}

\begin{table}[htbp]
  \centering
      \caption{{\footnotesize  The rank correlations of robot strengths, which are estimated by the WMPRC1 and WMPRC2 estimation procedures, of all robots on $M_{\ell}$ and $M_{\ell+1}$ matches, $\ell=6,\dots, 9,$ in the qualification stage.}}
      \label{tbl:table15}
\includegraphics[width=12cm,height=3.5cm]{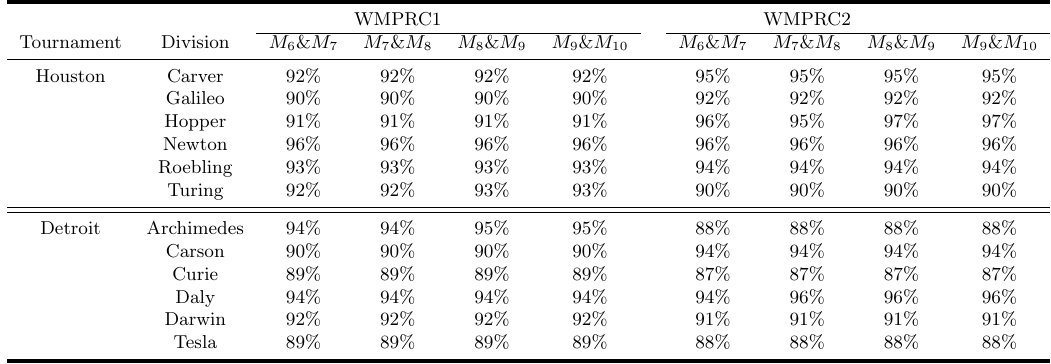}
\end{table}

\begin{table}[htbp]
  \centering
      \caption{{\footnotesize The rank correlations of robot strengths, which are estimated by the OPRC1 and OPRC2 estimation procedures, of model-based top-8 robots on $M_{\ell}$ and $M_{\ell+1}$ matches, $\ell=6,\dots, 9,$ in the qualification stage.}}
      \label{tbl:table16}
\includegraphics[width=12cm,height=3.5cm]{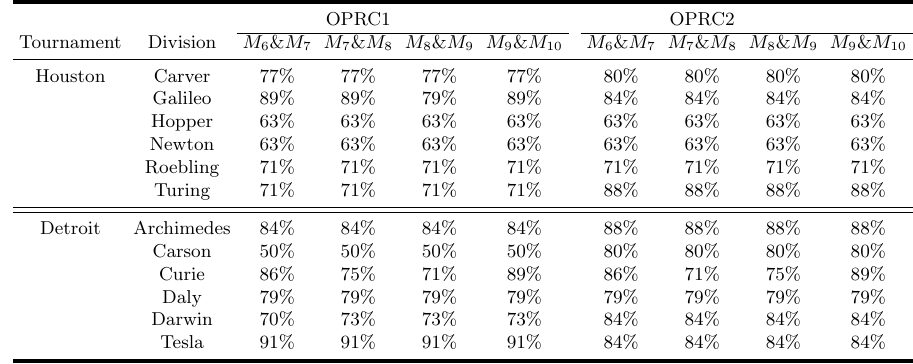}

\end{table}

\begin{table}[htbp]
  \centering
      \caption{{\footnotesize  The rank correlations of robot strengths, which are estimated by the WMPRC1 and WMPRC2 estimation procedures, of model-based top-8 robots on $M_{\ell}$ and $M_{\ell+1}$ matches, $\ell=6,\dots, 9,$ in the qualification stage.}}
      \label{tbl:table17}
\includegraphics[width=12cm,height=3.5cm]{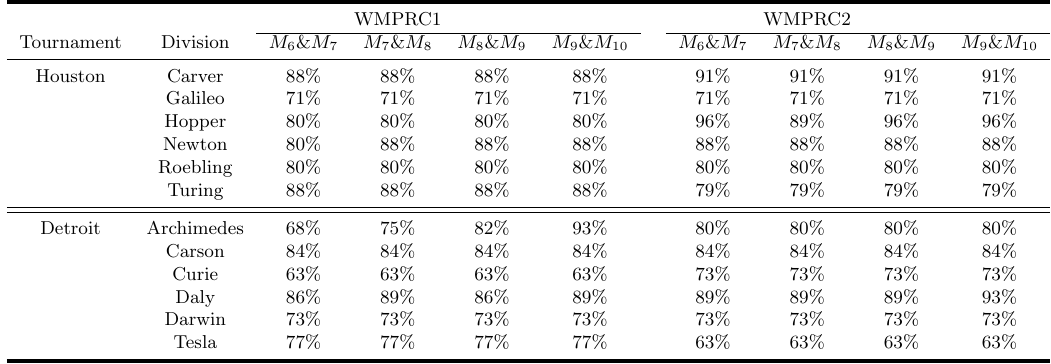}
\end{table}

\vspace{-0.05in}
\section{Conclusion and Discussion} \label{sec:conclusion}
In our application to the 2018 FRC championships, the OPR and WMPR models have poor predictive performance. This is mainly because  the ratio of the number of matches to the number of robots in each division is rather small and both models are highly over-parameterized. To enhance their predictive capacities, the OPRC and WMPRC models are proposed as possible avenues. In addition to generalize the model formulation of robot strengths, {we do not assume any particular distribution} on the underlying distributions of errors in the proposed models.
{In the analysis of such paired comparison data, the estimated accuracies of the WMPRC and OPRC models are about $14\%-26\%$ and $8\%-19\%$ higher than those of the WMPR and OPR models, respectively.}
It is notable that there is no need to take into account a specific-alliance advantage, a defensive component, and any {dynamic} effect in the WMPRC model.

{As stated in the introduction, performing well in the qualification stage comes with the benefit of being able to choose your alliance for the playoff stage. Compared to the other advantage of being guaranteed a spot in the playoff stage, this benefit is often taken for granted, with some teams opting to rely on chance qualification-round synergy, published OPR rankings, or cooperation history from tournaments or seasons past. Our methodology, shown to provide better predictive capabilities with regard to the playoff rounds, outputs a more reliable ranking of robots by rating that should assist in top teams' decision making during alliance selection. As a baseball field manager might look to past data and statistical models to guide a decision on whether or not to call in a relief pitcher or substitute a pinch hitter in a crucial inning, we hope that future FRC teams may look to our model to guide their endeavors at the highest levels of play, and that such adoption leads to a greater experience and appreciation of all facets of this complicated game.}

To sum up, our major contributions include:
\begin{enumerate}
\item Existing models are extended to more general models, which have better predictive performance in our application, with latent clusters of robot strengths;
\item Effective estimation procedures, in which the estimation problem is transferred to the model selection problem, are developed to simultaneously estimate the number of clusters, clusters of robots, and robot strengths;
\item A very flexible semiparametric regression model is proposed to predict a future match outcome; and
\item The stability of estimated robot strengths and accuracies is investigated to determine an appropriate number of matches in the qualification stage.
\end{enumerate}
Some measures are further used to assess the predictive ability of competing models and the agreement related to FRC ratings and model-based robot strengths. With slight modifications, our methodology should be successfully applied to other team games.

By taking into account random robot strengths, i.e. $\beta_{i}=\beta^{c_0}_{g_{i}}+\varepsilon_{g_{i}i}$
with $\varepsilon_{g_{i}i}$ having mean zero and variance 
$\sigma^{2}_{g_{i}}$, $i=1,\dots, K$, the proposed fixed effects model in
(\ref{eqn:linearcl}) can be further generalized to the following random effects model:
\begin{eqnarray}
Y=X^{c_0}\beta^{c_{0}}+
\bigg(\sum^{c_0}_{c=1}\sum^{K_c}_{j=1}\varepsilon_{cj}X_{(cj)}+\varepsilon\bigg), \label{eqn:linearran1}
\end{eqnarray}
where $X_{(cj)}$ is the designed covariate vector of $X_{(i)}$ with $g_{i}=c$, $i=1,\dots, K$, $\varepsilon_{cj}$,
which has mean zero and variance $\sigma^{2}_{c}$, $c=1,\dots,c_{0}$, $j=1,\dots,K_{c}$, are mutually independent, and $\varepsilon$ is independent of $\varepsilon_{cj}$'s. 
Obviously, the WMPRR model is a special case of the above model.
Different from the error in model (\ref{eqn:linearcl}), the error term $\big(\sum^{c_0}_{c=1}\sum^{K_c}_{j=1}\varepsilon_{cj}X_{(cj)}+\varepsilon\big)$ 
in model (\ref{eqn:linearran1})
is correlated. In particular, the model formulation 
avoids the problem of ties in $\beta_{i}$'s in model (\ref{eqn:linearcl}). It will be challenging to develop an appropriate estimation procedure for such a random effects model, especially for the determination of the number of clusters and clusters of robots. An investigation for its predictive ability would be worthwhile in future research. 
By making distributional assumptions on $\varepsilon_{cj}$'s and
$\varepsilon$, existing methods in the literature, such as \citet{laird:1982} and \citet{diggle:2002}, can also be used to estimate match-specific effects as well as robot-specific effects on each match score.

We note that no individual robot characteristic (e.g. the speed, weight, or height of a robot), which is expected to bring a contribution to alliance scores, is considered in existing models.  Another factor that affects match scores is penalty scores gained from opposing robots who violate game rules. For some FRC games, obstacles in the middle of the field change match-by-match relying on the audience selection. Since some robots might have an advantage on the designed obstacles in their matches, such an environmental factor needs to be carefully formulated.
The winning rate of each robot in its former matches should be also helpful in the model development.  As emphasized in this article, robots in the playoff stage are not randomly assigned to alliances and matches. A consideration of the above confounders is expected to enhance the predictive power of the current models. With this in mind, our goal is to continue improving the proposed model for the convenience of future robotics teams, leading to high quality competitions.

\section*{Acknowledgements} Chin-Tsang Chiang's research was partially supported by the Ministry of Science and Technology grant 109-2118-M-002-002-MY2 (Taiwan). Alejandro Lim completed a significant portion of the research work during his senior year at the Westminster Schools (Atlanta) and while visiting the Institute of Applied Mathematical Sciences, National Taiwan University as an intern.  The authors are grateful to Dr. Alvin Lim for useful discussions. 
\appendix
\section*{Appendix A}
\def\theequation{A.\arabic{equation}}
\setcounter{equation}{0}

Let $H^{c}=X^{c}\big(X^{c\top}X^{c}\big)^{-1}X^{c\top}$ and $\bar{H}^{c}=\bar{X}^{c}\big(\bar{X}^{c\top}\bar{X}^{c}\big)^{-1}\bar{X}^{c\top}$.
For the OPRC model, we derive that
\begin{eqnarray*}
\widehat{\beta}^{c}_{-s}&=&\widehat{\beta}^{c}-\frac{\big(X^{c\top}X^{c}\big)^{-1}
\big(\big(1-H^{c}_{M+s M+s}\big)e^{c}_{s}+H^{c}_{M+s s}e^{c}_{M+s}\big)X_{s}}
{\big(1-H^{c}_{s s}\big) \big(1-H^{c}_{M+s M+s}\big)-H^{c2}_{M+s s}}\nonumber\\
&&-\frac{\big(X^{c\top}X^{c}\big)^{-1}
\big(\big(1-H^{c}_{s s}\big)e^{c}_{M+s}+H^{c}_{M+s s}e^{c}_{s}\big)X_{M+s}}
{\big(1-H^{c}_{s s}\big) \big(1-H^{c}_{M+s M+s}\big)-H^{c2}_{M+s s}}, s=1,\dots, M.
\label{eqn:OPRC_cbhat}
\end{eqnarray*}
In light of the above result, $\widehat{F}^{c}_{-s}(v)$ has the form
\begin{eqnarray*}
\widehat{F}^{c}_{-s}(v)=\frac{1}{M-1}\sum_{s_{1}\neq s}I\big((Y_{M+s_{1}} - Y_{s_{1}}) - (X^{c}_{M+s_{1}} - X^{c}_{s_{1}})^\top \widehat{\beta}_{-s}^{c}\leq v\big)
\label{eqn:OPRC_F}
\end{eqnarray*}
with
\begin{eqnarray*}
&&(X^{c}_{M+s_{1}} - X^{c}_{s_{1}})^\top \widehat{\beta}^{c}_{-s}=(X^{c}_{M+s_{1}} - X^{c}_{s_{1}})^\top
\widehat{\beta}^{c}\nonumber\\
&&\hspace{0.3in}-
\frac{\big(H^{c}_{s M+s_{1}}-H^{c}_{s s_{1}}\big)
\big(\big(1-H^{c}_{M+s M+s}\big)e^{c}_{s}+H^{c}_{M+s s}e^{c}_{M+s}\big)}
{\big(1-H^{c}_{s s}\big) \big(1-H^{c}_{M+s M+s}\big)-H^{c2}_{M+s s}}\nonumber\\
&&\hspace{0.3in}
-\frac{\big(H^{c}_{M+s M+s_{1}}-H^{c}_{M+s s_{1}}\big)
\big(\big(1-H^{c}_{s s}\big)e^{c}_{M+s}+H^{c}_{M+s s}e^{c}_{s}\big)}
{\big(1-H^{c}_{s s}\big) \big(1-H^{c}_{M+s M+s}\big)-H^{c2}_{M+s s}}, s, s_{1}=1,\dots, M.
\label{eqn:OPRC_cpred}
\end{eqnarray*}
As for the WMPRC model, the properties in equation (8.5) of \citet{sen:1990} and Remark 4 enable us to have
\begin{equation*}
\widehat{\beta}^{*c}_{-s}=\widehat{\beta}^{*c}-\frac{\big(\bar{X}^{c\top}\bar{X}^{c}\big)^{-1}\bar{X}^{c}_{s}e^{c}_{s}}{\big(1-\bar{H}^{c}_{ss}\big)}
\end{equation*}
and
\begin{equation*}
  X^{c\top}_{s_1}  \widehat{\beta}^{c}_{-s}=\bar{X}^{c\top}_{s_1}
\widehat{\beta}^{*c}_{-s}, s,s_{1}=1,\dots, M.
\label{eqn:WMPRC_cbhat}
\end{equation*}
It follows that
\begin{equation*}
\widehat{F}^{c}_{-s}(v)=\frac{1}{M-1}\sum_{s_{1}\neq s}I\big(Y_{s_1} - X^{c\top}_{s_1} \widehat{\beta}_{-s}^{c}\leq v \big)
\end{equation*}
with
\begin{equation*}
X^{c\top}_{s_1}  \widehat{\beta}^{c}_{-s}=\bar{X}^{c\top}_{s_1}
\widehat{\beta}^{*c}
-\frac{\bar{H}^{c}_{s_{1}s} e^{c}_{s}}{\big(1-\bar{H}^{c}_{ss}\big)}, s, s_{1}=1,\dots, M.\label{eqn:WMPRC_F}
\end{equation*}

\bibliographystyle{cas-model2-names}

\bibliography{cas-sc}

\end{document}